\documentclass{article}
\usepackage[a4paper,left=2.5cm,right=2.5cm,top=2.5cm,bottom=2cm, bindingoffset=0mm]{geometry}
\usepackage{amsfonts}
\usepackage{dsfont}
\usepackage{mathtools}
\usepackage{bm}
\usepackage{amsthm}
\usepackage{hyperref}

\usepackage{multirow}
\usepackage{xcolor}
\definecolor{Gray}{gray}{0.93}
\usepackage{subcaption}
\usepackage{tikz}
\usetikzlibrary{shapes.geometric,arrows}
\tikzstyle{element} = [circle, minimum width=0.5cm, minimum height=0.5cm,text centered, draw=black, fill=Gray, text = black]
\tikzstyle{element2} = [circle, minimum width=0.5cm, minimum height=0.5cm,text centered, draw=black, fill=Gray!90!black, text = black]
\tikzstyle{element3} = [circle, minimum width=0.5cm, minimum height=0.5cm,text centered, draw=black!50!white, fill=white, text = white]
\tikzstyle{roundarrow} = [thick,->,>=stealth]

\newcommand{\RM}[1]{\mathrm{\MakeUppercase{\romannumeral #1}}}

\title{Multi player Parrondo games with rigid coupling}
\author{Sandro Breuer\footnote{sandrobreuer@gmail.com} ~and
  Andreas Mielke\footnote{mielke@tphys.uni-heidelberg.de (corresponding author)}
  \\
  Institut für Theoretische Physik\\
  Ruprecht-Karls-Universität Heidelberg\\
  Philosophenweg 12\\
  D-69121 Heidelberg, Germany}
\date{\today}

\begin{document}

\maketitle

\begin{abstract}
  In the original Parrondo game, a single player combines two losing strategies
  to a winning strategy. In this paper we investigate the question
  what happens, if two or more players play Parrondo games in a
  coordinated way. We introduce a strong coupling between the players
  such that the gain or loss of all players in one round is the same.
  We investigate two possible realizations of such a coupling.
  For both we show that the coupling increases the gain per player.
  The dependency of the gain on the various parameters of the games
  is determined. The coupling can not only lead to a larger gain,
  but it can also dominate the
  driving mechanism of the uncoupled games. Which driving mechanism
  dominates, depends on the type of coupling. Both couplings are set
  side by side and the main similarities and differences are
  emphasised.

  Keywords: Noise induced transport; Parrondo’s paradox; Markov chains;
  multiplayer Parrondo games; collective coupling effect
\end{abstract}

\section{Introduction}

A Parrondo game \cite{Harmer_Abbott_1999,Abbott_Harmer_1999} consists of two simple
games, typically realized by flipping biased coins, which are played
in some regular or randomly alternating sequence. The interesting
effect occurring here is that even if the two games lead to a
systematic loss if played sufficiently long, the combination yields a
systematic win.

Originally, Parrondo invented these games to illustrate the occurrence
of noise induced transport in so called Brownian motors, for a review
see \cite{Reimann_2002}. Brownian motors have been proposed first by Magnasco
\cite{Magnasco1993} as a model for intra-cellular transport
created by motor proteins like kinesin, which move along
micro-tubuli. In the simplest form, a Brownian motor is a Brownian
particle moving in a time-dependent periodic potential without inversion
symmetry. The Brownian particle is driven by a white noise process
as usual, representing a finite temperature of the system.
The time dependence of the potential can be either periodic,
see {\emph e.g.} \cite{Reimann_2002} or stochastic. In the
case of a stochastic additive or multiplicative noise added to the
potential, this additional noise process must have a finite correlation
time, but can be otherwise an arbitrary noise process, see {\emph e.g.}
\cite{Mielke_1995} for the additive noise and \cite{Mielke_1995_2}
for the multiplicative noise process.
The combination of
the broken inversion symmetry and the additional additional
periodic or stochastic time dependence yields a 
non-vanishing stationary current. The Parrondo games can be viewed as
a discretized version of Brownian motors \cite{Allison_Abbott_2002}.
But the study of Parrondo games is an interesting topic independently of that initial
motivation. Not only are Parrondo games probably the most
simple systems where the coupling of systems with detailed balance
yields a new system where detailed balance is broken and therefore a
stationary current occurs, but they also can have direct link to living systems. Lai and Cheong \cite{Cheong2020} for example connect the Parrondo games with  "societal ideas
of redistribution, cooperation, voting, performance,
and resource growth to bring about 'winning'
outcomes in a social group." \cite{Cheong2020} Furthermore, Cheong \emph{et al.} \cite{Cheong2022} investigate the winning strategy in bacteriophages because of Parrondo's Paradox.

Motivated by the fact that motor proteins like kinesin have two heads
which couple to the micro-tubuli, Klumpp \emph{et al.} \cite{Klumpp_2001} investigated
the noise-induced transport of two strongly coupled particles. They
showed that the transport of a system of two strongly coupled
particles is more efficient than the transport by a single
particle. This motivates us to study coupled multi-player Parrondo
games.  The idea is to introduce two or more coupled players, each
playing the same Parrondo game.  To our knowledge, such multi-player
Parrondo games with correlated, interacting players
have not been studied so far.  As in
\cite{Klumpp_2001}, we concentrate on strong coupling where in each
step all players win or lose. The aim is to investigate, how such a
coupling can be realized, whether in such a coupled system the gain or
loss is higher than in the uncoupled system, and if due to the
coupling new mechanisms for the creation of a stationary current
occur, which are not present for a single player.

Our paper is organized as follows. We first introduce and review
ordinary Parrondo games, mainly to fix the notation. In
Sect. \ref{coupledGames} we introduce two different couplings for two
or more players.  These coupled systems can be easily investigated by
simulations or by methods using discrete-time Markov chains (DTMC) to
calculate the stationary current. We discuss the results and
investigate the mechanisms that yield to the higher stationary current
in the case of coupled players. Finally, in Sect. \ref{Outlook} we
summarize our results, give an outlook and propose some future
research in this area.

\section{Parrondo games}\label{parrondoGames}

Let us first introduce some basics about usual Parrondo games, mainly
to fix the notation and to introduce the methods of DTMC used to
investigate the capital current. First, we state the original
definition of the Parrondo games given by Abbott and Harmer
\cite{Harmer_Abbott_1999,Abbott_Harmer_1999}. Let $x_0 \in \mathbb{N}_0$ be the
initial capital of a player. The player can win or lose one capital
unit $\Delta x = \pm 1$ in every round (negative capital should be
possible) and hence obtains the capital $x(n) \in \mathbb{Z}$ after
$n$ rounds. The winning probability $p$ depends on the choice of game
and is periodic as a function of capital with period $M$. Game~A is a
homogeneous process (the same coin is used in every round), game~B is
capital dependent (one of two coins is chosen depending on the capital).
Parrondo originally chose the probabilities given in table
\ref{parrondoGames_probs}. In general, game C is the result of periodically or
randomly switching between games A and B. 

\renewcommand{\arraystretch}{1.4}
\begin{table}
  \centering
  \begin{tabular}{|c|c|c|c|c|}
    \hline
    &  \textbf{capital} & \textbf{game~A} & \textbf{game~B} & \textbf{game C (random)}\\
    \hline
    $p_0$  & $x \bmod 3 = 0$ & \multirow{2}{*}{$\frac{1}{2} - \epsilon $}& $\frac{1}{10} - \epsilon $ & $q_0\cdot\frac{1}{2}+q_1\cdot\frac{1}{10} - \epsilon$\\
    $p_1$ & $x \bmod  3 \neq 0$ & &$\frac{3}{4} - \epsilon$ & $q_0\cdot\frac{1}{2}+q_1\cdot\frac{3}{4} - \epsilon$\\
    \hline
  \end{tabular}
  \caption{Winning probabilities $p_{0,1}$ for the original Parrondo
    games~\cite{Abbott_Harmer_1999}. The variable $q_{0,1}$ defines the probability
    to choose game A,B, respectively, $\epsilon$ is the so called bias
    parameter. Parrondo originally chose $\epsilon = 0.005$ and
    $M=3$.}
  \label{parrondoGames_probs}
\end{table}
\renewcommand{\arraystretch}{1}

We consider only random switching.
We implement that via the dichotomous random process $z \in \{0,1\}$
indicating the choice of game A and B, respectively, at the beginning
of every round with the probabilities
\begin{align}
  P(\mathrm{play \ A}) = P(z=0) & \eqqcolon q_0 \, ,\\
  P(\mathrm{play \ B}) = P(z=1) & \eqqcolon q_1 = 1-q_0 \, .
\end{align}
The variable $p_{0,1}^{\mathrm{A,B,C}}$ denotes the winning probability for the
games A, B and C for $(0)$ a capital multiple of $M$ and $(1)$
otherwise. 

Since the transition probabilities are periodic functions of the
capital, we can restrict the discussion to the reduced state space
$\mathbb{Z}/M\mathbb{Z}$ and choose periodic boundary
conditions. Indeed, since the winning probabilities only depend on the
current capital and hence on the current state, Parrondo games are
DTMC. The transition matrix in the reduced state space
$Q=(Q_{ij})_{i,j\in \mathbb{Z}/M\mathbb{Z}}$ with 
\begin{equation}\label{Markov-property}
  Q_{ij}(n) \coloneqq P(x_{n+1}=i|x_n=j)
\end{equation}
is finite and time-homogeneous and has the explicit form
\cite{Abbott_Harmer_2002},
\begin{equation}\label{transition matrix}
  Q(n) = Q = \left(\begin{array}{ccccc}
                     0        & 1-p_1 &  &  & p_1\\
                     p_0  & 0         &\ddots&&\\
                              & p_1    & \ddots &1-p_1&\\
                              &               &\ddots&0&1-p_1\\
                     1-p_0       &    &  &p_1 & 0
       
                   \end{array}\right) \, .
               \end{equation}

Let $\Bar{x}=x \bmod M$ be the reduced capital and $P$
be the probability distribution on the reduced state
space, i.e. $P_i$ be the probability for
$\Bar{x}=i$. We then obtain for the time evolution
\begin{equation}\label{time evolution}
  P_i(n+1) = \sum_{j=0}^{M-1} Q_{ij}P_j(n) \Longleftrightarrow P(n+1) = QP(n) \, .
\end{equation}
Especially, the stationary distribution $\pi$ of a
time-homogeneous DTMC is given by $\pi = Q\pi$.

Under certain conditions, DTMC converge
towards such a stationary distribution
\cite{Markov_Chains}, p. 150. In fact, one can show that
the DTMC of the original Parrondo games converges
against a unique stationary distribution in the sense
of
$\underset{n\rightarrow \infty}{\mathrm{lim}} \big|
P_i(n) - \pi_i \big| = 0$ for odd $M$.
This is true if the games are \textit{irreducible},
\textit{aperiodic}, and \textit{positive recurrent},
see \cite{Markov_Chains}, p. 118ff, 150.

Since every
state communicates with every other one by simply winning or losing
$D$ times if the capital difference is $D$, the DTMC of the games is
\textit{irreducible} and consists of only one communicating class
(the mutual communication induces an equivalence relation)
\cite{Markov_Chains}, p. 80.

Furthermore, the period of state $i$
\cite{Markov_Chains}, p. 84.
\begin{equation}\label{period}
  d_i = \mathrm{gcd}\left\{n\geq1: P(x_n = i|x_0 = i) > 0\right\} \, ,
\end{equation}
which is the same for communicating states and hence can be defined
for the whole class, becomes $1$ for an odd $M$. This is obvious
since the capital can return to its initial value after two rounds
and after winning $M$ times in a row. Therefore, the games with an
odd $M$ are \textit{aperiodic}.

The last property is
\textit{positive recurrence}, which is defined
by~\cite{Markov_Chains}, p. 111
\begin{equation}\label{positive recurrence}
  P(\tau_i<\infty) = 1 \land E(\tau_i) < \infty \, ,
\end{equation}
where we used the first recurrence time
$\tau_i \coloneqq \mathrm{min} \{n\geq1: x_n = i | x_0 =
i\}$. Brémaud \cite{Markov_Chains}, p. 122 proves that an irreducible,
time-homogeneous DTMC with a finite state space is positive
recurrent, hence positive recurrence is shown for the reduced
Parrondo games with an odd $M$.  For time-homogeneous, ergodic
(irreducible, aperiodic and positive recurrent) DTMC with a finite
state space, there is not only an unique stationary distribution
\cite{Markov_Chains}, p. 118ff, but also a convergence theorem
stating the given convergence \cite{Markov_Chains}, p. 150. Since all
properties apply to the reduced Parrondo games with odd $M$, the
convergence is proven.

We now know that there is a unique probability distribution against
which every initial distribution converges in the long-run limit. From
this we can compute a stationary capital current giving us the
long-run capital difference per round. Indeed, since the stationary
current is independent of the initial distribution, we can use it to
evaluate and compare different games. This strategy is important: As can be seen for game B, a game does not have to be a
martingale but nevertheless might be fair in the long-run limit corresponding to a
vanishing capital current \cite{Abbott_Harmer_1999, Abbott_Harmer_Parrondo_Taylor_2001}. The reason for this is that a martingale must be balanced at every single step. This is obviously not the case for game B. However, the game can be balanced on average anyway. This unusual effect is described in more detail in \cite{Costa2005}.

Let
\begin{equation}\label{winning probability}
  p^{\mathrm{win}}_{j} = \bigg\{\begin{array}{ll}
                         Q_{j+1,j} & \mathrm{|\ \ }j\neq M-1\\
                         Q_{0, M-1}& \mathrm{|\ \ }j=M-1
                       \end{array}
\end{equation}
be the winning probabilities.                     
The stationary current can be obtained from the stationary
distribution $\pi$. Toral \emph{et al.}  \cite{Toral_Amengual_Mangioni_2003_2} derive,
using the corresponding master equation, a discrete form for the
probability current, which reduces for the Parrondo games to
\cite{Toral_Amengual_Mangioni_2003_2}
\begin{equation}
  {J_i (n) = -(1-p^{\mathrm{win}}_{i})P_i(n) + p^{\mathrm{win}}_{i-1}P_{i-1}(n)} \, .
\end{equation}
Summing over all states and considering periodic boundary conditions
($p^{\mathrm{win}}_{-1} = p^{\mathrm{win}}_{M-1},\ P_{-1} = P_{M-1}$), this gives
\begin{align}\label{capital current Parrondo}
  J   &= \sum_{i=0}^{M-1} J_i = \sum_{i=0}^{M-1} \left[
        -(1-p^{\mathrm{win}}_i)\pi_i + p^{\mathrm{win}}_i \pi_i \right] \notag\\
      &= \sum_{i=0}^{M-1} 2p^{\mathrm{win}}_i\pi_i - 1 = 2E(p^{\mathrm{win}}) - 1
\end{align}
in the stationary case.

In particular we observe that $J=0$ is equivalent to
$E(p^{\mathrm{win}}) = \frac{1}{2}$ which is the definition of fairness given
by Abbott and Harmer \cite{Abbott_Harmer_2002}. Indeed, for the original Parrondo
games with $M=3$ it is possible to compute the stationary distribution
analytically and therefore obtain an analytic expression for the
stationary capital current which leads to
\begin{equation}\label{current M=3}
  J = \frac{3 \left(2 p_{0} p_{1}^{2} - 2 p_{0} p_{1} + p_{0} - p_{1}^{2} + 2 p_{1} - 1\right)}{2 p_{0} p_{1} - p_{0} + p_{1}^{2} - 2 p_{1} + 3} \, .
\end{equation}
Inserting the probabilities from table \ref{parrondoGames_probs} with
$q_0 = 0.5$, one can easily see that there is a range of $\epsilon$
for which $J_A, J_B < 0$ but $J_C > 0$ leading to Parrondo's
Paradox. Hence, the formalism for investigating the Parrondo games
with one player is simple: Given the transition matrix for a DTMC, we
obtain the stationary distribution by computing the eigenvector for
the eigenvalue $1$ and with equations \ref{winning probability} and
\ref{capital current Parrondo} calculate the capital current for the
games. It has to be noted that this does not only apply to the
original Parrondo games but to every transition matrix corresponding
to a one-dimensional DTMC with states sorted by their capital.

For the coupled Parrondo games we want to compare the capital current
for different parameters, especially the period $M$ and the width of
the barrier $d$ which is given by the number of capitals within one
period corresponding to a small winning probability in game B. For the
original Parrondo games, one can see in table
\ref{parrondoGames_probs} that $d=1$. In order to only investigate the
coupling effect, the uncoupled games should lead to the same current
when varying the parameters. Therefore, we modify the winning
probabilities accordingly. Since game $A$ is independent of the
capital, it does not change. However, the probabilities for game B
have to be adapted so that it is always fair and game C with the
random combination for $q_0 = 0.5$ induces a current $J=J_{\mathrm{const}}$ for
all parameters. This is achieved numerically by introducing the
probabilities as variables and computing the roots of the function
$[J_B, J_C - J_{\mathrm{const}}]$. We implemented the numerics in Python using
the package \textit{scipy.optimize.fsolve}. We choose those roots
satisfying $p_0 \in (0, 0.5)$ and $p_1 \in (0.5, 1)$. For the
variation of $M$ and $d$ we choose $J_{\mathrm{const}} = 0.05$ and
$J_{\mathrm{const}} = 0.02$, respectively. The results are listed in tables
\ref{parrondo games varying M} and \ref{parrondo games M=19 varying
  d}.

\begin{table}
  \centering
  \begin{tabular}{|c|c|c|c|c|}
    \hline
    &  \textbf{capital} $\bm{x}$ & $\bm{M = 3}$ & $\bm{M = 5}$ &$\bm{M = 7}$ \rule[-7pt]{0pt}{0pt}\rule{0pt}{14pt}\\
    \hline
    ${p_0}$  & $x \bmod M = 0$ & 0.04200574 & 0.03467694 & 0.02181704\\
    ${p_1}$ & $x \bmod  M$ $\neq 0$ & 0.82685756 & 0.69669249 & 0.65335836\\
    \hline
    & $\bm{M = 11}$ &$\bm{M = 19}$ & $\bm{M = 29}$ & $\bm{M = 49}$ \rule[-7pt]{0pt}{0pt}\rule{0pt}{14pt}\\
    \hline
    ${p_0}$ & 0.00849192 &0.00135921 & 1.45686849$\cdot 10^{-4}$ & 1.88929755$\cdot 10^{-6}$\\
    ${p_1}$ & 0.61680554 &0.59064647 & 0.57822635 & 0.56821420\\
    \hline
  \end{tabular}
  \caption{Adapted winning probabilities $p_{0,1}$ for game B for
    different $M$ with $\epsilon = 0$. For $\epsilon \neq 0$,
    $\epsilon$ has to be subtracted. Game~A does not change,
    $p_0 = p_1 = 0.5 - \epsilon$.}
  \label{parrondo games varying M}
\end{table}

\begin{table}
  \centering
  \begin{tabular}{|c|c|c|c|c|c|}
    \hline
    &  \textbf{capital} $\bm{x}$ & $\bm{d = 1}$ & $\bm{d = 2}$ & $\bm{d = 3}$ & $\bm{d=4}$ \rule[-7pt]{0pt}{0pt}\rule{0pt}{14pt}\\
    \hline
    ${p_0}$  & $x \bmod M < d$ & 0.02213766 & 0.04204007 & 0.03145708 & 0.00312982\\
    ${p_1}$ & $x \bmod  M \geq d$& 0.55241902 & 0.59092394 & 0.65533964 & 0.85097252\\
    \hline

  \end{tabular}
  \caption{Adapted winning probabilities $p_{0,1}$ for game B for
    different $d$ with $\epsilon = 0$ and $M=19$. For
    $\epsilon \neq 0$, $\epsilon$ has to be subtracted. Game~A does
    not change, $p_0 = p_1 = 0.5 - \epsilon$.}
  \label{parrondo games M=19 varying d}
\end{table}

With these probabilities and $\epsilon = 0.5 \cdot 10^{-6}$, game A
and B are losing games for all periods and widths of the barrier,
respectively. However, game C with $q_0 = 0.5$ is a winning game, we
obtain Parrondo's Paradox. For $\epsilon \neq 0$ there is a slight
deviation of the current $J_{\mathrm{const}}$ for the uncoupled games. However,
considering the change caused by the coupling, this is irrelevant for
small bias parameters.

\section{The coupled Parrondo games}\label{coupledGames}

We are now in the position to introduce coupled Parrondo games. The
idea is to have more than one player and to couple the players so that
they do not play independently.  In this paper we only investigate the
rigid coupling where in each round all players win or lose the same
amount.  We first consider two players $\RM{1}$ and $\RM{2}$.  We
denote their capital as $x_{\RM{1}, \RM{2}} \in \mathbb{Z}$. They win
or lose one capital unit in every round. The individual winning
probabilities are defined by the Parrondo games for a single
player. However, the players are not independent. The probability
${P(x_{\RM{1}}(n+1) = i,x_{\RM{2}}(n+1) = k\ |\ x_{\RM{1}}(n) =
  j,x_{\RM{2}}(n) = l)}$ is obtained by combining the individual
probabilities according to the couplings, which will be explained in
the following.

As above, we consider the state space $\mathbb{Z}/M\mathbb{Z}$. Since
we will show that the Parrondo games with rigid coupling can be
reduced to the ordinary Parrondo games with modified transition
probabilities, they will also have a stationary distribution in this
state space. In order to find a transition matrix, we define a
projection between both capitals and a single variable

\begin{equation}\label{Projektion}
  \begin{array}{ccc}
    (i,k) \in \{\mathbb{Z}/M\mathbb{Z}\}^2 & \longleftrightarrow & a \in \{\mathbb{Z}/M^2\mathbb{Z}\}\\
    \bar{x}_{\RM{1}} = i \ , \  \bar{x}_{\RM{2}} = k & \longleftrightarrow & \bar{x}_{\RM{1}} = \lfloor a/M \rfloor \ , \ \bar{x}_{\RM{2}} = a \bmod M
  \end{array}
\end{equation}
with $\bar{x}_{\RM{1}, \RM{2}} \coloneqq x_{\RM{1}, \RM{2}} \bmod M$
and the floor function $\lfloor x \rfloor$.  Especially, the variable
$a$ is obtained by $a = M\cdot i + k $. We will always use this
transformation for $a \leftrightarrow i,k$ and
$b \leftrightarrow j,l$. The transition matrix then becomes:
\begin{equation}\label{kombinierte Übergangswahrscheinlichkeit}
  Q_{ab} = P(\bar{x}_{\RM{1}}(n+1) = i, \ \bar{x}_{\RM{2}}(n+1) = k \ \ | \ \ \bar{x}_{\RM{1}}(n) = j, \ \bar{x}_{\RM{2}}(n) = l) \, .
\end{equation}
For calculating this matrix, we need to introduce the coupling between
both players. In the following we define
two couplings that combine both players' winning or losing
probabilities for a collective gain or loss.

The first approach is the double-play coupling.
Both players play the individual Parrondo games and have to win
or lose at the same time for a collective gain or loss. Additionally,
all the other cases are excluded by setting their transition
probabilities to zero and renormalizing the others. This coupling is
motivated by Brownian motion of two coupled particles: In an
infinitesimal time interval, both particles simultaneously have to
move to the right or to the left.

The second approach is the single-play coupling. One of the players is
chosen at the beginning of every round with a certain probability and
plays the individual Parrondo games. The result will then also be
applied to the other player. This coupling is motivated by certain
motor proteins: The heads of a motor protein are coupled, but can
attach and detach independently during the ATP hydrolysis. This
corresponds to the random choice of one player.

We adapt the notation of the original Parrondo games by adding the
super- or subscripts $\RM{1}$, $\RM{2}$, e.g. $p_{\RM{1}}^X$ is the
winning probability of player $\RM{1}$ when playing game
$X \in \{A,B\}$, which is of course capital dependent itself. The
individual winning and losing probabilities are obtained from the
single Parrondo games in table \ref{parrondo games varying
  M}. However, when varying the width of the barrier, we use the
probabilities in table \ref{parrondo games M=19 varying d}.  Since the
external force is the same for both particles in the continuous case,
we always set $\epsilon_{\RM{1}} = \epsilon_{\RM{2}}$.

With this we obtain the combined winning probabilities for the
double-play games as
\begin{equation}\label{multiplied winning probabilities double-play
    games}
  p^{XY} = \frac{p_{\RM{1}}^Xp_{\RM{2}}^Y}{p_{\RM{1}}^Xp_{\RM{2}}^Y + (1-p_{\RM{1}}^X)(1-p_{\RM{2}}^Y)}
\end{equation}
when player $\RM{1}, \RM{2}$ plays game $X, Y \in \{A, B\}$,
respectively. Let $Q^{XY}$ be the corresponding transition matrix. The
entries for possible transitions result from the collective winning
and losing probabilities similar to equation \ref{winning probability}
(note that they depend on the current capitals of the players!), the
remaining entries are set to zero. By combining the different
possibilities for the games we obtain
\begin{equation}\label{transition matrix double-play games}
  Q = q^{\RM{1}}_0q^{\RM{2}}_0 \cdot Q^{AA} + q^{\RM{1}}_0q^{\RM{2}}_1 \cdot Q^{AB} + q^{\RM{1}}_1q^{\RM{2}}_0 \cdot Q^{BA} + q^{\RM{1}}_1q^{\RM{2}}_1 \cdot Q^{BB} \, .
\end{equation}
For the single-play coupling we define the probability $p_{\RM{1}}$ of
choosing player $\RM{1}$ and $p_{\RM{2}} = 1-p_{\RM{1}}$. Since the
second player always wins or loses simultaneously, the collective
winning probability is
\begin{equation}\label{winning prbability single-play games}
  p^{\mathrm{win}, \mathrm{win}} = p_{\RM{1}}p_{\RM{1}}^{\mathrm{win}} + p_{\RM{2}}p_{\RM{2}}^{\mathrm{win}} = p_{\RM{1}}q^{\RM{1}}_0p^A_{\RM{1}} + p_{\RM{1}}q^{\RM{1}}_1p^B_{\RM{1}} + p_{\RM{2}}q^{\RM{2}}_0p^A_{\RM{2}} + p_{\RM{2}}q^{\RM{2}}_1p^B_{\RM{2}} \, .
\end{equation}
Here $p_{\RM{1}}^{\mathrm{win}}$ is the single winning probability for the
random alternation between game A and B
for player $\RM{1}$. The
same applies to the transition matrices.

We now demonstrate that the Parrondo games with rigid coupling can be
reduced to the original Parrondo games with modified transition
probabilities. For the case of two players, one state is defined
either by both capitals or by one capital and the capital
difference. Since rigid coupling induces constant capital differences,
the state is only determined by the capital of one player when a
certain initial condition is given. Hence, we can separate the Markov
chain of the coupled Parrondo games for different capital
differences. This is due to the reducibility of the Markov chain. The
different states with the same capital difference form an irreducible
equivalence class which is equivalent to the single Parrondo games
but with modified transition probabilities because of the coupling. In
fact, considering the capital differences
$D = (\bar{x}_{\RM{2}} - \bar{x}_{\RM{1}}) \mod M \in \{0, ..., M-1\}$
(convention: $\bar{x}_{\RM{2}} > \bar{x}_{\RM{1}}$), the transition
matrix of the coupled games has the form
($Q_{{D}} \in \mathbb{R}^{M\times M}, \ D \in \{0,\ ...,\ M-1\} $, all
other entries vanish)
\begin{equation}\label{Nach Äquivalenzklassen umsortierte
    Übergangsmatrix}
  Q^\prime \  = \left(
    \begin{array}{cccc}
      Q_{{0}} & & \\
              & \ddots & \\
              & & Q_{{M-1}}
    \end{array}\right) \, .
\end{equation}
                 
It is obvious that $Q_{{D}}$ is the transition matrix of the class
${D}$ and with the corresponding probability distribution $P_{{D}}(n)$
at step $n$ we can write
\begin{equation}\label{Zeitentwicklung für eine Klasse}
  P_{{D}}(n+1) = Q_{{D}}P_{{D}}(n) \, .
\end{equation}

For the different equivalence classes and hence the different capital
differences we can then use the existence of a stationary probability
distribution in order to determine the stationary capital current. For
an initial probability distribution that contains different capital
differences with nonzero probability, the different solutions for the
stationary capital current have to be added with the corresponding
stochastic weights.

Hence we have already found a formalism for investigating the coupled
games: The transition matrices are calculated according to the
different couplings and dependent on certain parameters such as the
period, noise parameters or width of the barrier. They can then be
reduced to the transition matrices of the equivalence classes for
which we can compute the stationary distribution and probability
current.

It may be mentioned that we can always treat the probability current
and capital current as equivalent since the probability current
reflects the change of states and the discrete capital difference is
set to 1. Especially, since the winning of both players only
corresponds to a single state change, the capital current in the
coupled case can be directly compared to the capital current in the
single case.

We also want to investigate the Parrondo games with multiple ($N>2$)
players coupled. Therefore we have to slightly modify the
formalism. The state is now given by all capitals,
$x = (x_\RM{1}, x_{\RM{2}}, ...) \in \{\mathbb{Z}/M\mathbb{Z}\}^N$. It
may be interesting to mention that one state is again determined by
the capital of one player and the capital differences between two
consecutive players in a certain order. Since the equivalence classes
are determined by the capital differences due to rigid coupling, we
have again reduced the multiple-coupled Parrondo games to the single
Parrondo games with modified transition probabilities. For the
double-play coupling, all winning (losing) probabilities are
multiplied and renormalized as well as stochastically weighted with
the probabilities for choosing the different games $A$ and $B$ for
each player. For the single-play coupling, the individual transition
probabilities are stochastically weighted with the probabilities
$p_{\RM{1}, \RM{2}, ...}$ of choosing the players
$\RM{1}, \RM{2}, ...$ at the beginning of every round.

One can once more define a projection onto one variable. The resulting
transition matrix can then be reduced to the various submatrices of
the equivalence classes corresponding to different capital differences
for which we can compute the stationary capital current. However,
since the projection does not effect the stationary capital current,
we will not go into detail here.

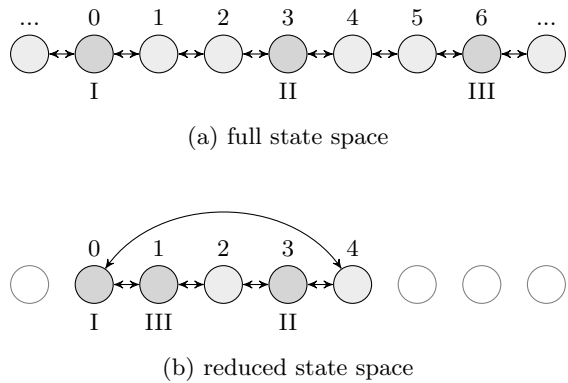
\begin{figure}[ht]
  \centering
  \begin{subfigure}[b]{0.5\linewidth}
    \centering
    \begin{tikzpicture}[<->,>=stealth',auto,node distance=0.85cm,
      thin]
      \node[label={\small ...}] (0) [element] {}; \node[label={\small
        0}, label=below:\small$\RM{1}$] (1) [element2, right of=0] {};
      \node[label={\small 1}] (2) [element, right of=1] {};
      \node[label={\small 2}] (3) [element, right of=2] {};
      \node[label={\small 3}, label=below:\small$\RM{2}$] (4)
      [element2, right of=3] {}; \node[label={\small 4}] (5) [element,
      right of=4] {}; \node[label={\small 5}] (6) [element, right
      of=5] {}; \node[label={\small 6}, label=below:\small$\RM{3}$]
      (7) [element2, right of=6] {}; \node[label={\small ...}]  (8)
      [element, right of=7] {}; \path[every
      node/.style={font=\sffamily\small}] (0) edge node [right] {} (1)
      (1) edge node [right] {} (2) (2) edge node [right] {} (3) (3)
      edge node [right] {} (4) (4) edge node [right] {} (5) (5) edge
      node [right] {} (6) (6) edge node [right] {} (7) (7) edge node
      [right] {} (8);
    \end{tikzpicture}
    \caption{full state space}
  \end{subfigure}
  \hfill
  \begin{subfigure}[b]{0.5\linewidth}
    \centering \vspace{0.5cm}
    \begin{tikzpicture}[<->,>=stealth',auto,node distance=0.85cm,
      thin]
      \node (0) [element3] {}; \node[label={\small 0},
      label=below:\small$\RM{1}$] (1) [element2, right of=0] {};
      \node[label={\small 1}, label=below:\small$\RM{3}$] (2)
      [element2, right of=1] {}; \node[label={\small 2}] (3) [element,
      right of=2] {}; \node[label={\small 3},
      label=below:\small$\RM{2}$] (4) [element2, right of=3] {};
      \node[label={\small 4}] (5) [element, right of=4] {}; \node (6)
      [element3, right of=5] {}; \node (7) [element3, right of=6] {};
      \node (8) [element3, right of=7] {}; \path[every
      node/.style={font=\sffamily\small}] (1) edge node [right] {} (2)
      (2) edge node [right] {} (3) (3) edge node [right] {} (4) (4)
      edge node [right] {} (5) (5) edge[bend right = 55] node [left]
      {} (1);
    \end{tikzpicture}
    \caption{reduced state space}
  \end{subfigure}
  \caption[Illustration of the multiplayer state space]{Illustration
    of the multiplayer state space. Here the state for $M=5$ and $N=3$
    as well as $D=3$ between two consecutive players and
    $x_{\RM{1}}=0$ is shown. The circles are states, the numbers above
    the circles reflect the capital and the roman numbers underneath
    the circles reflect the players in a certain order.}
  \label{multiplayer illustration}
\end{figure}

The multiple-player state space is illustrated in figure
\ref{multiplayer illustration} for $M=5$, $N=3$, and $D=3$
between two consecutive players. In the reduced state space, the
capitals of the first and third player are only one capital unit
apart. This effect will be important when interpreting the results for
multiple players below.

\section{The stationary current of multi-player games}

\subsection{Double-play games}

\begin{figure}[ht]
  \centering
  \includegraphics[width=0.5\linewidth]{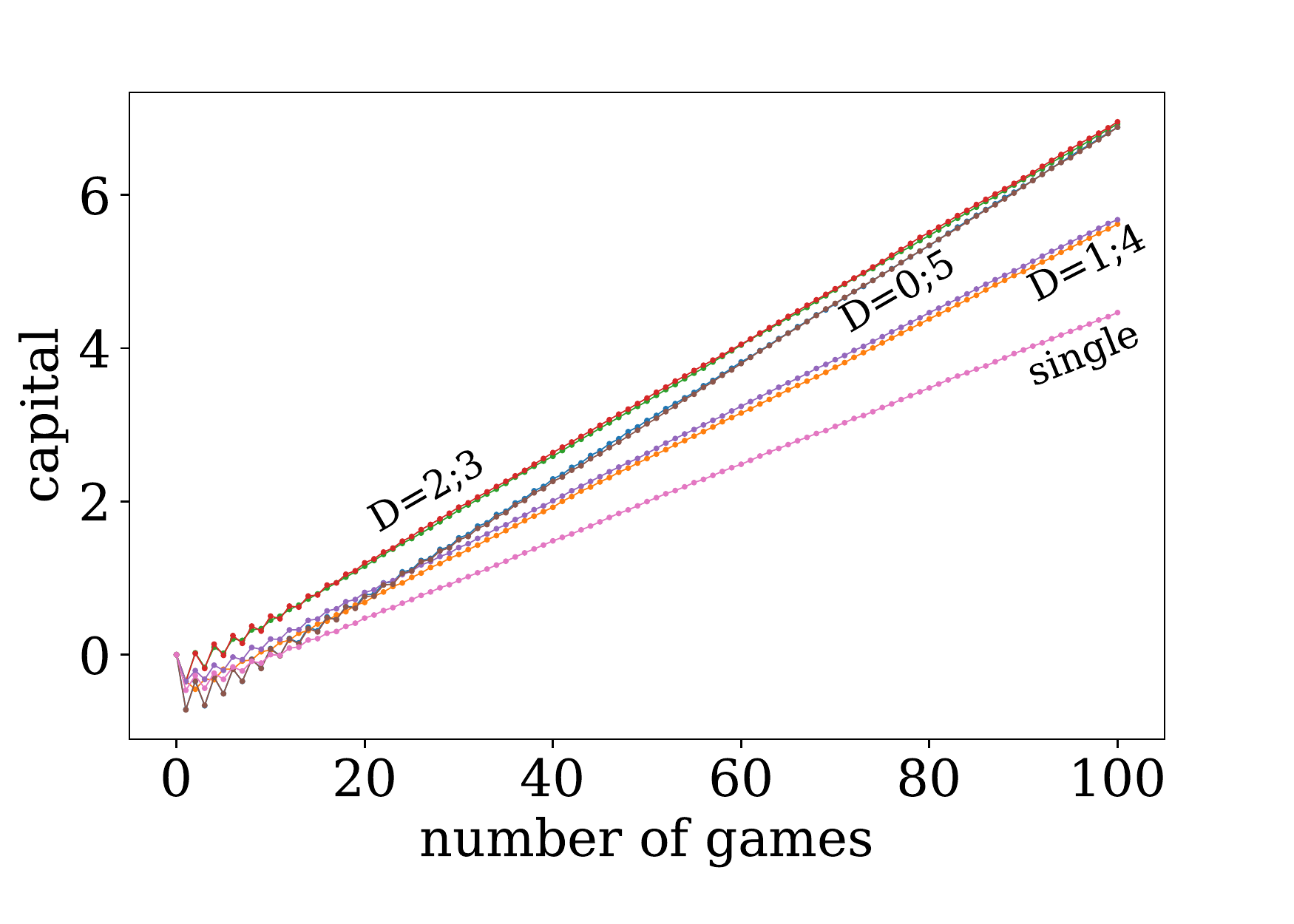}
  \caption[Simulation of the double-play games]{Simulation of the
    double-play games averaged over 50000 repetitions for
    $\epsilon_{\RM{1}} = \epsilon_{\RM{2}} = 0.5 \cdot 10^{-6}$, $M=5$
    and $q_0^{\RM{1}} = q_0^{\RM{2}} = 0.5$ compared to the uncoupled
    (single) case with $q_0 = 0.5$. The capital of player $\RM{1}$ is
    shown.}
  \label{Simulation double}
\end{figure}

We first analyse the double-play games. As a first step, we simulate the
games. We consider the capital flow of the different classes and
therefore always choose initial conditions that only contain states of
the same class. The simulation is done for $M=5$, the results are
shown in figure \ref{Simulation double}.

It can be seen that during the first rounds, some capital is gained
and lost alternately. This is due to the initial capital
$\bar{x}_{\RM{1}} = 0$. After several rounds, the distribution
converges against the stationary one and the capital gradient
approaches a positive constant.

Additionally, the slopes of $D$ and $M-D$ are similar, respectively,
even though one can observe slight offsets which are the result of the
initial condition.

The next step is to analyse the games with methods of DTMC and to
compare the results with the uncoupled Parrondo games. In particular,
the dependence on the capital difference $D$ and the noise parameters
$q^{\RM{1}, \RM{2}}_0$ is examined, the width of the barrier and the
number of players are varied.

\begin{figure}[ht]
  \centering
  \includegraphics[width=0.5\linewidth]{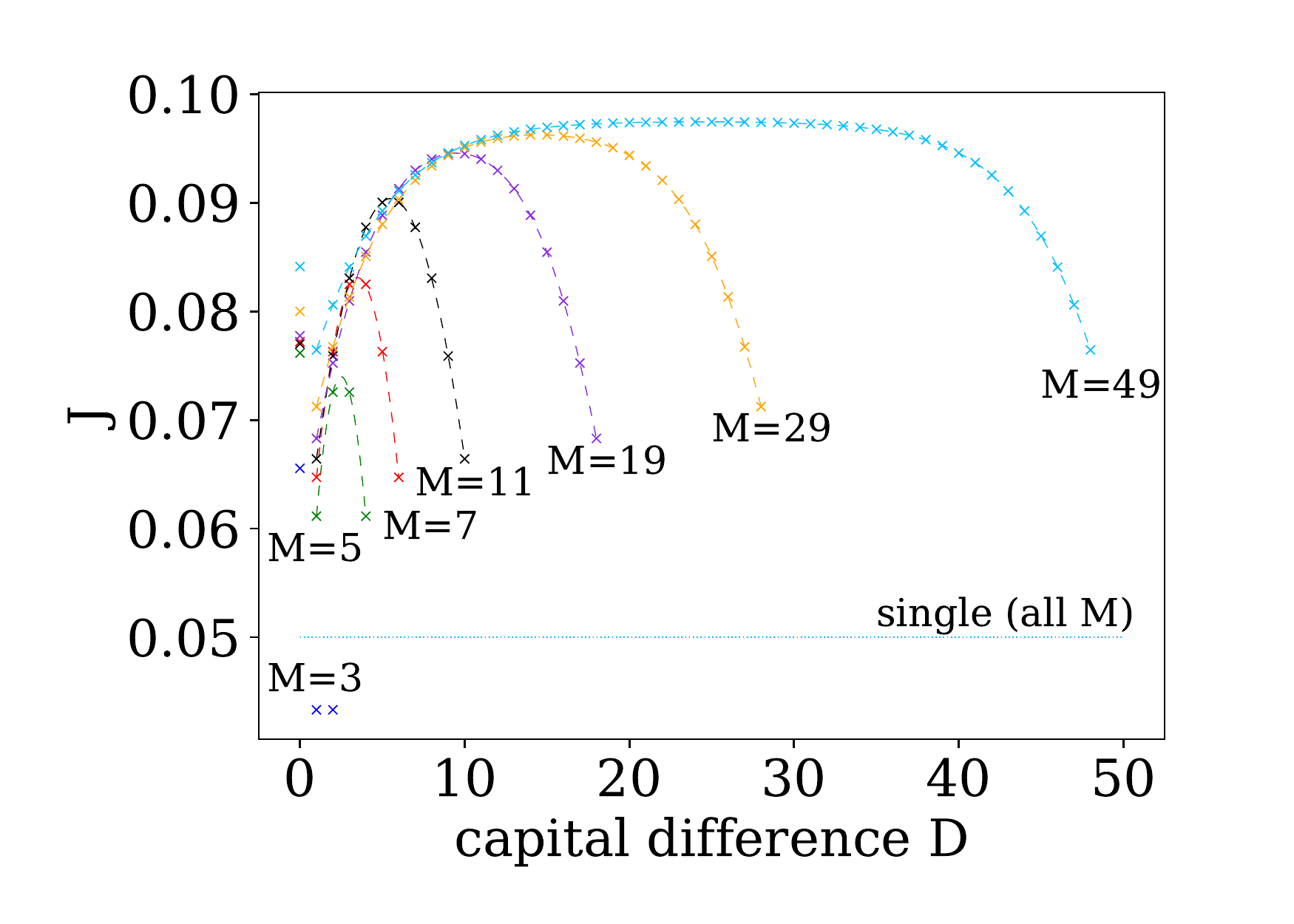}
  \caption[Capital current of the double-play games for different
  periods]{Capital current of the double-play games for different
    periods. We choose
    $\epsilon_{\RM{1}} = \epsilon_{\RM{2}} = 0.5 \cdot 10^{-6}$,
    $q^{\RM{1}}_0 = q^{\RM{2}}_0 = 0.5$ and $q_0 = 0.5$ for the
    uncoupled (single) case. We fit a parabola for
    $M \in \{5, 7, 11\}$ and a polynomial of degree $M-2$ for
    $M \in \{19, 29, 49\}$, $D=0$ is not taken into account.}
  \label{J(D)_double_M}
\end{figure}

First, we change the capital difference for different
periods. We choose the same parameters as for the simulation
\ref{Simulation double}. The results are portrayed in figure
\ref{J(D)_double_M}. It is obvious that the capital current in the
uncoupled case is the same for all periods, this is due to the
modification of the probabilities accordingly.

The long-run current in the simulation can be estimated by a linear
fit after several rounds. The results obtained that way coincide with
the stationary current calculated using DTMC.

It can be observed that the capital current of the uncoupled games
takes the value $J = 0.05$ up to a small deviation due to
$\epsilon \neq 0$. Moreover, the assumption is verified that the
capital currents of $D$ and $M-D$ are equal. Therefore, we could show
the accordance of the simulation and computation of the stationary
capital current. In figure \ref{J(D)_double_M} we can observe the
following:

\textbf{1a) The current for ${M > 3}$ is always larger than in the
  uncoupled case.}  In the uncoupled case the capital accumulates in
front of $\bar{x} = 0$ when playing game $B$ because the winning
probability at $\bar{x} = 0$ is smaller (a small or big winning
probability is in the following always referred to game B) than
$0.5$, but the winning probability before is larger than $0.5$. This
is a kind of barrier. The switch between games $A$ and $B$ allows the
capital to cross the barrier more likely. When two players are coupled
with a positive capital difference, at least one player has a big
winning probability and hence helps the other player to cross the
barrier since both winning probabilities are multiplied. This is the
driving mechanism of the double-play games.

However, the period $M=3$ does not match with this observation. Here
the capital current for a positive capital difference is smaller than
in the uncoupled case. A possible explanation could be that for $M=3$
and $D>0$ there are more states with at least one player having a
small winning probability than states with no player having a small
winning probability at all. The same argument leads to the result that
the capital current of $D=0$ is always larger than for
$D=1$. Interestingly, the coupling with $D=0$ has a vanishing extent
but nonetheless leads to an increase in the capital current compared
to the uncoupled case. This is the result of nonlinear effects within
the multiplication of the winning probabilities. We will later see
that this occurs because the width of the barrier is chosen to be
1. When this width is enlarged, small capital differences can not have
any positive effect on the capital current.

\textbf{1b) The current has a maximum around ${M/2}$ for positive
  capital differences.}  For ${D < \lfloor M/2 \rfloor}$ and
$D > \lceil M/2 \rceil$ the current decreases symmetrically. For large
periods some kind of saturation current is reached and the maximum of
the former dependence flattens. For small periods parabolas are fitted
and amazingly agree with the data, for larger periods polynomials of
degree $M-2$ are fitted in order to clarify the curve.  The symmetry
of the curve is a result of the choice of parameters: Since the
parameters are the same for both players, the players are
indistinguishable and the long-run behaviour of the games is the same
for $D$ and $M-D$ (the convention of the direction of the coupling
was only important for the definition of the classes but does not
effect the winning probabilities).

Particularly interesting is the maximum at
$D \in \{\lfloor M/2 \rfloor, \lceil M/2 \rceil\}$. In these cases,
the capitals are distributed as widely as possible over the state
space considering the periodic boundary conditions. When the first
player is located at a barrier ($\bar{x} = 0$), the second player can
help him cross the barrier as efficiently as possible. This
observation has also been found in the continuous case analysed by
Klumpp \emph{et al.} \cite{Klumpp_2001}. They chose the
asymmetry of the potential the other way around and therefore the
probability current is negative in their case.

\textbf{1c) The maximum increases with ${M}$.}  This is a result of
the coupling. In the last observation we deduced that the capital
current is maximal for a wide distribution of capitals over the state
space. A larger period increases this effect and hence the driving
mechanism. However, there is some kind of saturation effect for large
periods.

\begin{figure}[ht]
  \centering
  \includegraphics[width=0.5\linewidth]{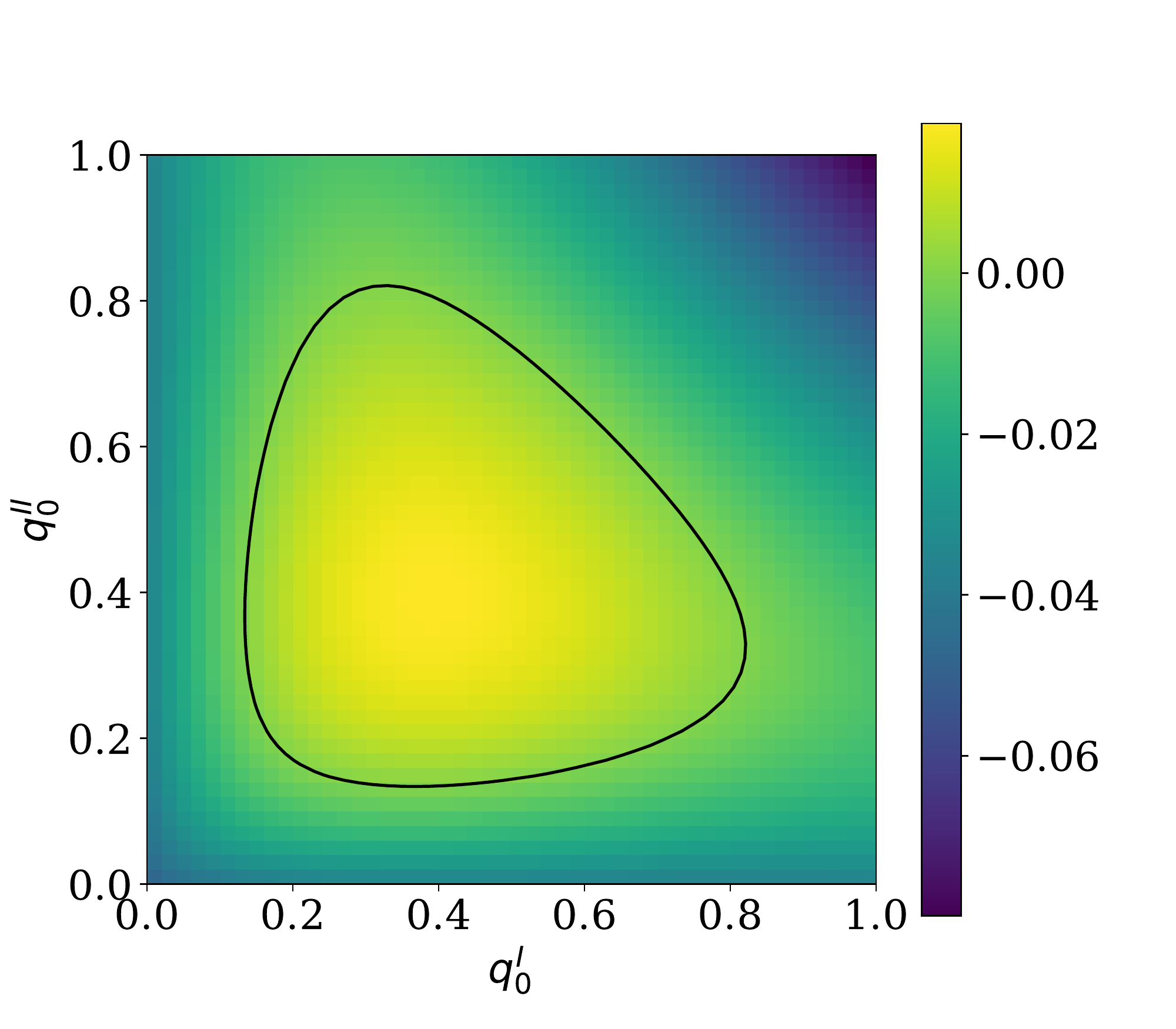}
  \caption[Capital current of the double-play games as a function of
  $q_0^{\RM{1}, \RM{2}}$]{Capital current of the double-play games as
    a function of $q_0^{\RM{1}, \RM{2}}$ for $M=7$,
    $\epsilon_{\RM{1}, \RM{2}} = 0.02$ and $D = 3$. The black line
    shows the contour line for $J = 0$.}
  \label{J(q)_double_M7_eps002}
\end{figure}

We now analyse the dependence on the noise parameters. First, we
compute the capital current as a function of the noise parameters
$q_0^{\RM{1}}$ and $q_0^{\RM{2}}$. As an example, we choose $M=7$,
$\epsilon_{\RM{1}} = \epsilon_{\RM{2}} = 0.02$ and $D=3$. The results
are shown in figure \ref{J(q)_double_M7_eps002}. 

We observe the
following characteristics, which hold for other values of $M$ and $D$
as well:

\textbf{1d) For sufficiently large values of ${\epsilon}$, the area in
  which the current is positive is finite.}  This is one of the
essential results and tells us that the switch between games $A$ and
$B$ is important for the double-play games: This switch is the only
noise process in the system and is therefore essential for the
games and for crossing the barrier.

\textbf{1e) For ${\epsilon \rightarrow 0}$ the area of positive
  current enlarges and reaches the boundaries
  ${q}_0^{{\RM{1}},{\RM{2}}}{\in \{0,1\}}$.}  This was observed when
varying the bias parameter for the current as a function of the noise
parameters but is not displayed here. The expansion of the area is a
result of the increasing winning probabilities for
$\epsilon \rightarrow 0$. It is quite interesting that a positive
current can be achieved even though one player always plays the same
game.

\textbf{1f) The four double-deterministic points in the corners are
  the last ones to reach a positive current for a shrinking bias
  parameter.}  This confirms the result in observation \textbf{1d)}:
When there are no noise processes and both players always play the
same game, respectively, the capital current is minimal. However, it
can be shown that all points but $q^{\RM{1}}_0 = q^{\RM{2}}_0 = 1$ can
attain a positive current. This is due to the homogeneous
probabilities of game A which, for a positive bias parameter, can
never lead to a winning game.

\begin{figure}[ht]
  \centering
  \includegraphics[width=0.5\linewidth]{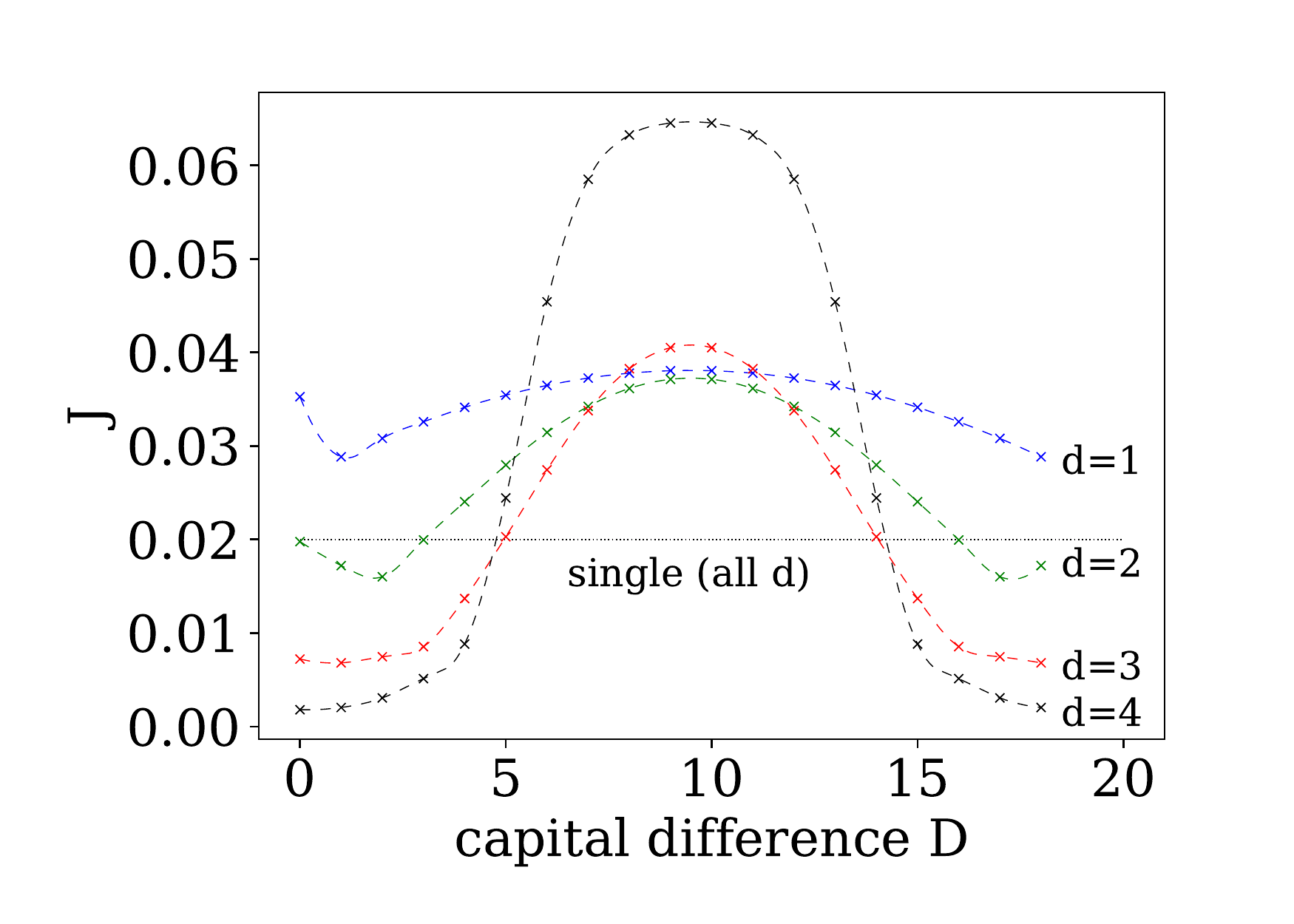}
  \caption[Capital current of the double-play games for different
  widths of the barrier]{Capital current of the double-play games for
    different widths of the barrier $d$ for $M=19$,
    $q^{\RM{1}}_0 = q^{\RM{2}}_0 = 0.5$,
    $\epsilon_{\RM{1}} = \epsilon_{\RM{2}} = 0.5 \cdot 10^{-6}$ and
    $q_0 = 0.5$ for the uncoupled (single) case. The curve is
    clarified with a spline interpolation of second degree.}
  \label{J(D)_double_M19_d}
\end{figure}

Another interesting observation can be made when varying the width of
the barrier. Therefore we choose the probabilities in table
\ref{parrondo games M=19 varying d} as well as
$q^{\RM{1}}_0 = q^{\RM{2}}_0 = 0.5$ and
$\epsilon_{\RM{1}} = \epsilon_{\RM{2}} = 0.5 \cdot 10^{-6}$. The
results are portrayed in figure \ref{J(D)_double_M19_d}. 

It is obvious
that observation \textbf{1b)} applies. Nevertheless, there is a change
in the characteristics at the edges of the plot:

\textbf{1g) For ${d>1}$ the capital current at the edge of the curve
  is smaller than in the uncoupled case and the slope of the curve
  flattens.}  A minimum distance between the capitals is required so
that the coupling can have a constructive effect. This is the same
observation as in~\cite{Klumpp_2001}. Klumpp \emph{et al.} \cite{Klumpp_2001}
argue that the coupling is a positive driving mechanism only if the
equilibrium distance between the particles is larger than the
potential barrier. Then one particle can help the other crossing the
barrier as can be seen in \cite{Klumpp_2001}, Fig. 4. This
quantitative conclusion can not be deduced in our discrete case. For
example, the capital current for $d=3$ does not exceed the current of
the uncoupled case until $D=5$.

\begin{figure}[ht]
  \centering
  \includegraphics[width=0.5\textwidth]{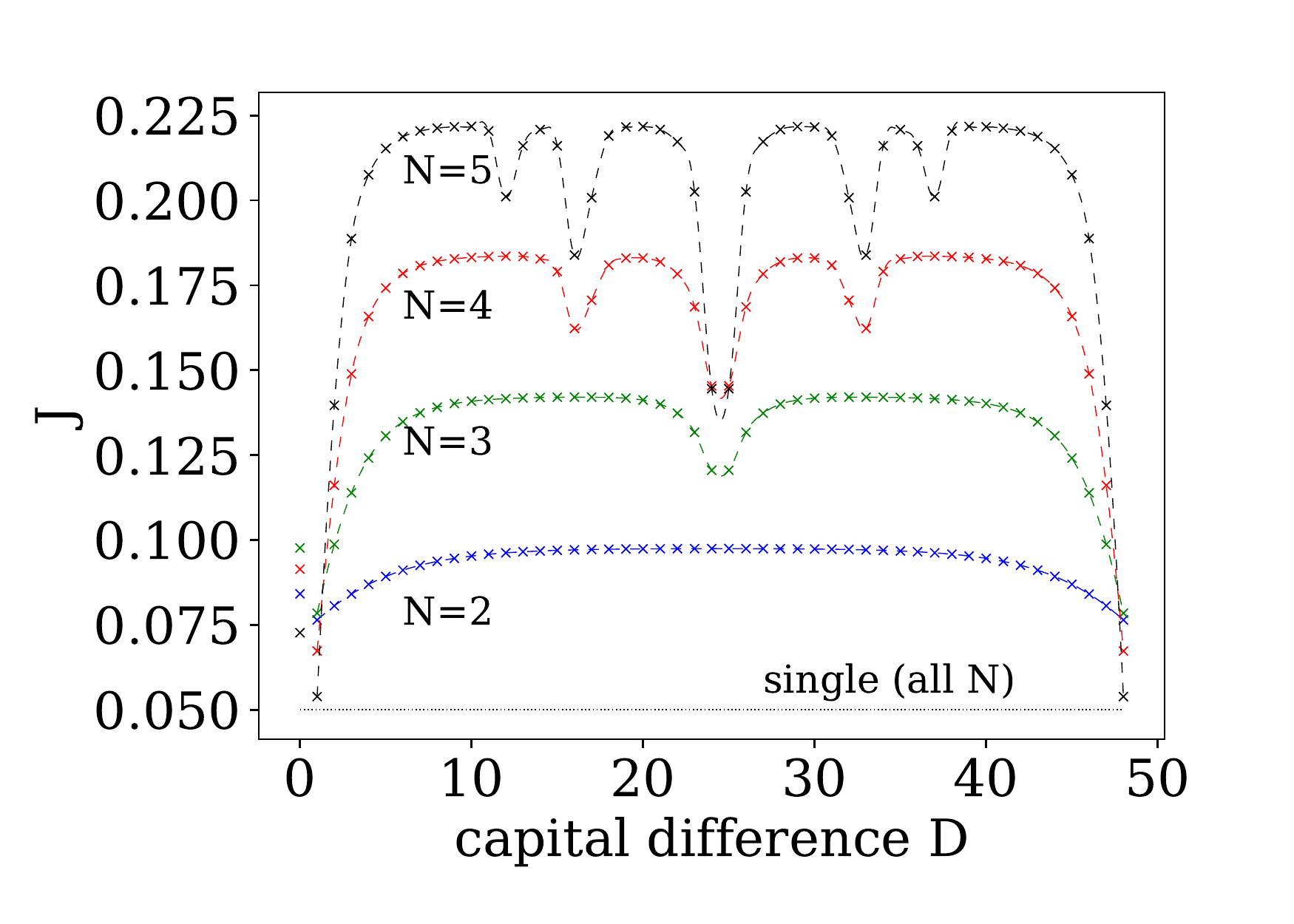}
  \caption[Capital current of the double-play games for a varying
  number of players]{Capital current of the double-play games for a
    varying number of players $N$ for $M=49$,
    $\epsilon = 0.5 \cdot 10^{-6}$ and $q_0 = 0.5$ for all players and
    in the uncoupled (single) case. The curve is clarified with a
    spline interpolation of second degree.}
  \label{J(D)_double_multiplayer}
\end{figure}

Up to now we analysed the behaviour of the capital current as a
function of different parameters for two players. We now want to
consider more than two players. We restrict the investigation to
equidistant capitals, hence the capital difference $D_{i,i+1}$ between
two consecutive players $i$ and $i+1$ in a certain order is a
constant,
$D_{i,i+1} = D \in \{0, ..., M-1\} \ \forall i \in \{1, ...,
N-1\}$. Since we will always choose the same parameters for all
players, the exact order is irrelevant. The results for $M=49$ are
illustrated in figure \ref{J(D)_double_multiplayer}. 

Of course many of
the observations for two players can also be found here. Particularly
important is the behaviour dependent on the number of players:

\textbf{1h) For ${M>7}$ the number of extreme points increases with
  ${N}$ or stays constant.} Especially, one can observe that the
current is maximal when the capitals are distributed as widely as
possible over the state space and minimal for multiple capitals being
close together and hence contributes to the conclusion of observation
\textbf{1b)}. We illustrate this taking $N=4$ as example: The capital
current increases until $D=12$. If the first player has a capital
multiple of M, the fourth player has a capital difference of
$D_{1,4} = 3D = 36$ to the first one and reaches a multiple of $M$
after 13 other capital units. For an increasing $D$ and taking into
account the periodic boundary conditions, the fourth capital
approaches the first one and is as near as possible for $D=16$, the
current is minimal. This repeats several times dependent on $M$ and
$D$. The effect is also illustrated in figure \ref{multiplayer
  illustration}.

\textbf{1i) For ${D=M/2}$ there is a minimum for ${N>2}$ and a maximum
  for ${N=2}$.}  This is a direct result of observation \textbf{1h)}:
The capitals are as widely distributed as possible for $N=2$ and
concentrated at two points for $N>2$, the current is maximal and
minimal, respectively.

It may be mentioned that these effects are only that clear because $M$
is chosen quite large. For smaller periods, the effects of the
discretization are visible. However, we do not want to go into detail
here.

\subsection{Single-play games}

We now investigate the single-play games in a similar manner.
As before we simulate the games
and subsequently analyse them with methods of DTMC so that we can
compare the results with the uncoupled case.

\begin{figure}[ht]
  \centering
  \includegraphics[width=0.5\linewidth]{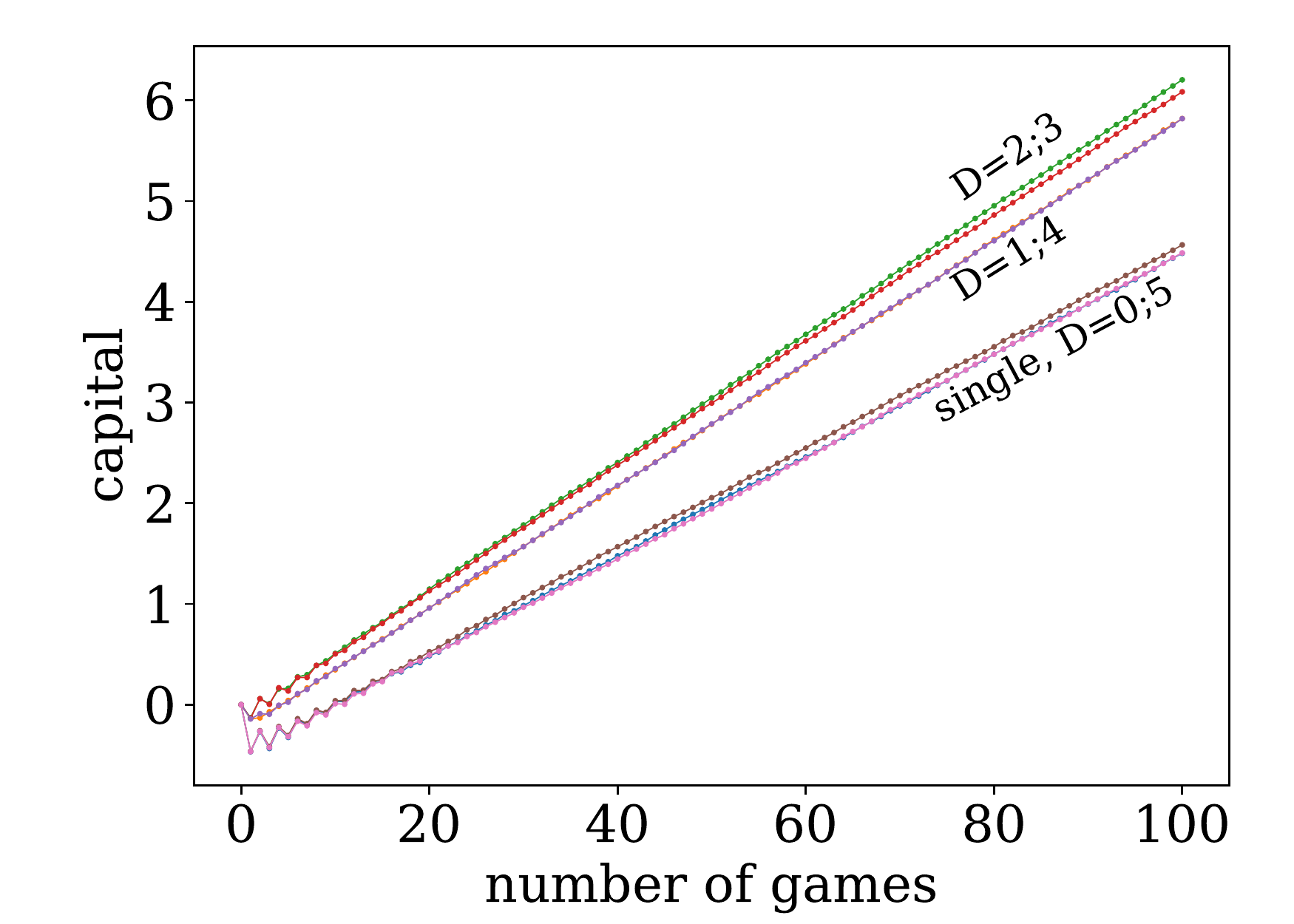}
  \caption[Simulation of the single-play games]{Simulation of the
    single-play games averaged over 50000 repetitions for
    $\epsilon_{\RM{1}} = \epsilon_{\RM{2}} = 0.5 \cdot 10^{-6}$,
    $M=5$, $q_0^{\RM{1}} = q_0^{\RM{2}} = 0.5$ and $p_{\RM{1}} = 0.5$
    compared to the uncoupled game with $q_0 = 0.5$. The capital of
    player $\RM{1}$ is shown.}
  \label{Simulation single}
\end{figure}

For the simulation we choose initial distributions only containing
elements of one class, the result is portrayed in figure
\ref{Simulation single}.

The simulation shows the same behaviour as in the double-play
games. The fluctuation during the first rounds is a result of the
initial condition and we observe the convergence against a stationary
capital current. Moreover, the capital differences $M$ and $M-D$
produce the same capital current in the long-time limit, small offsets
can be explained by the initial conditions.

\begin{figure}[ht]
  \centering
  \includegraphics[width=0.5\linewidth]{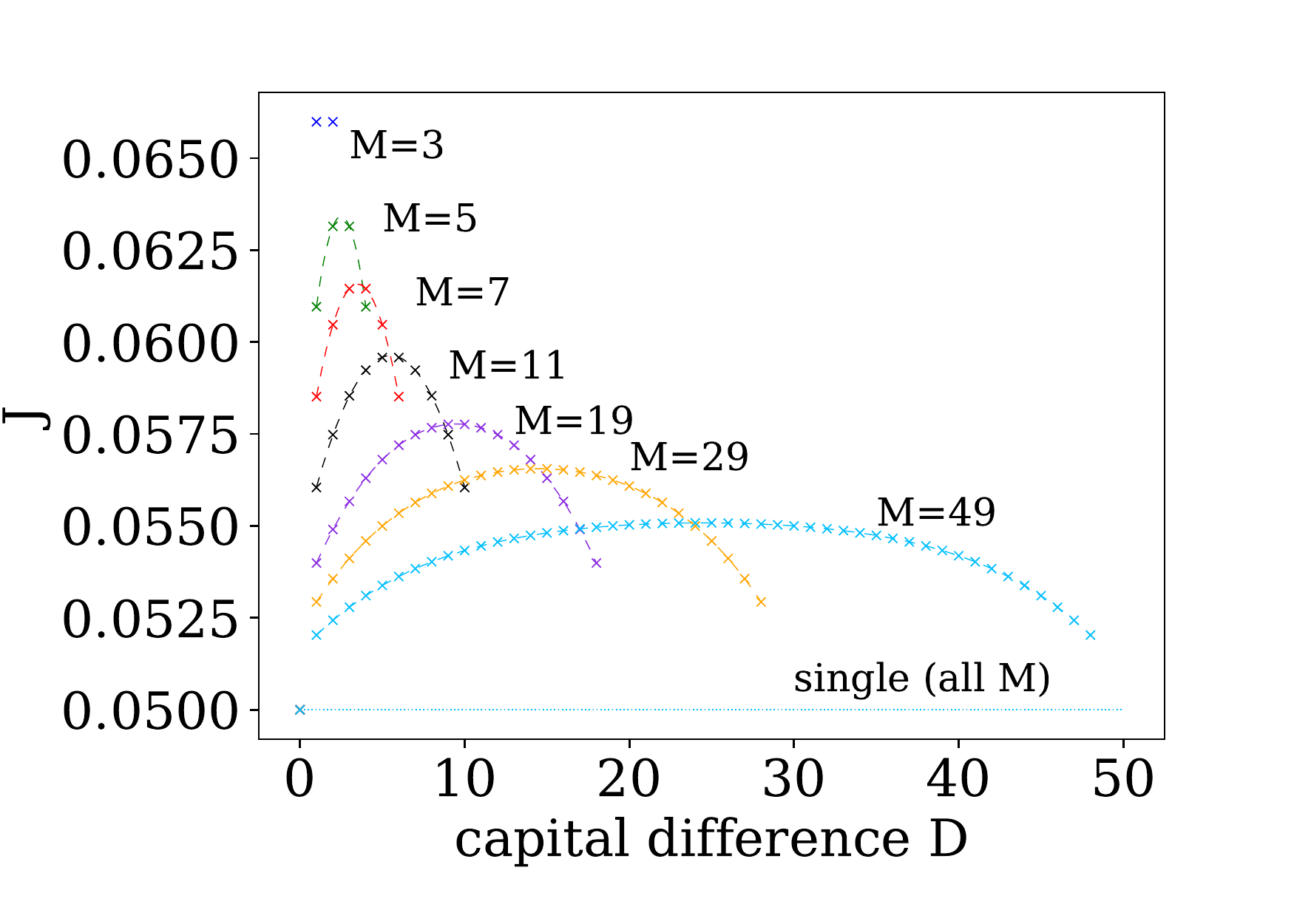}
  \caption[Capital current of the single-play games for different
  periods]{Capital current of the double-play games for different
    periods. We choose
    $\epsilon_{\RM{1}} = \epsilon_{\RM{2}} = 0.5 \cdot 10^{-6}$,
    $q^{\RM{1}}_0 = q^{\RM{2}}_0 = 0.5$, $p_{\RM{1}} = 0.5$ and
    $q_0 = 0.5$ for the uncoupled case. We fit a parabola for
    $M \in \{5, 7, 11\}$ and a polynomial of degree $M-2$ for
    $M \in \{19, 29, 49\}$, $D=0$ is not taken into account (here all
    data points overlap).}
  \label{J(D)_single_M}
\end{figure}

As before, we vary different parameters. First, we can again compute the capital
current for different periods and capital differences. We choose the
same parameters as in the simulation. The results are shown in figure
\ref{J(D)_single_M}.  
  
The simulation agrees with the computation via
DTMC.  We can again verify that the differences $D$ and $M-D$ show the
same current. Figure \ref{J(D)_single_M} gives us the following
insights:

\textbf{2a) The current is larger than in the uncoupled case.}  Hence
the coupling has a positive effect on the capital current. However,
the explanation differs from the double-play coupling: In the
double-play games both winning probabilities are multiplied and
therefore the extra driving mechanism is that both players always
determine the winning probability together. The single-play coupling
introduces a new noise process, the choice of the active player in
each round. For this reason, even if one of the players is at a
barrier, there is a chance that the other player is chosen and helps
him crossing the barrier. This is the new driving mechanism of the
single-play games.

Again, $D=0$ is a special case: Since all parameters are chosen to be
the same for both players, this is then equivalent to the individual
Parrondo games as can be observed in figure \ref{J(D)_single_M} (all
points overlap at that point).

\textbf{2b) The current has a maximum around ${M/2}$ for positive
  capital differences.}  This observation is the same as observation
\textbf{1b)} and can be explained equally since the effect does not
depend on the coupling but on the capital distribution over the state
space.

\textbf{2c) The maximum decreases with ${M}$.}  Apparently, the
driving mechanism of the single-play games has a different impact for
varying periods and is, in contrast to the double-play games, more
efficient the closer the barriers are.

\begin{figure}[ht]
  \centering
  \includegraphics[width=0.5\linewidth]{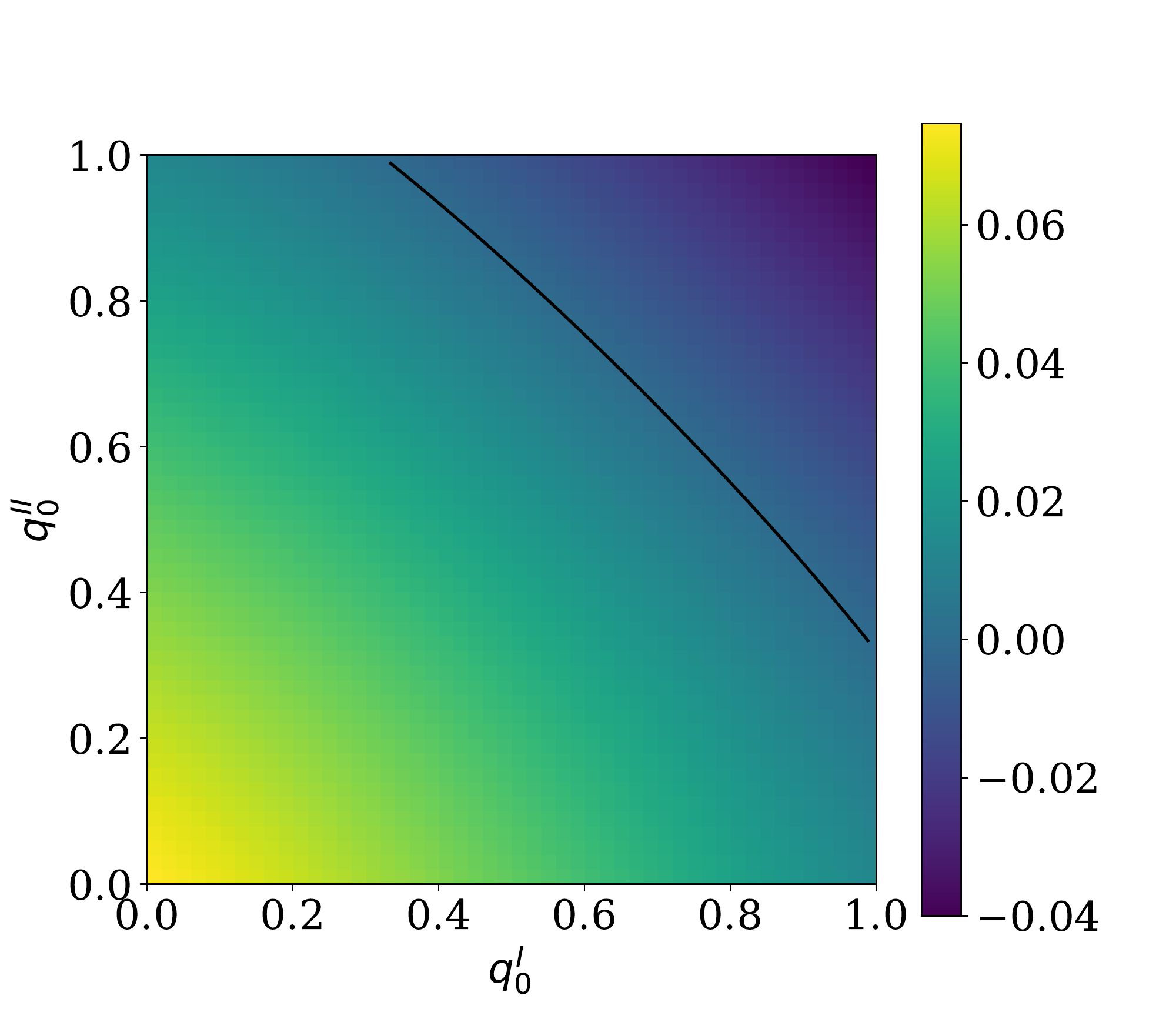}
  \caption[Capital current of the single-play games as a function of
  $q_0^{\RM{1}, \RM{2}}$]{Capital current of the single-play games as
    a function of $q_0^{\RM{1}, \RM{2}}$ for $M=7$,
    $\epsilon_{\RM{1}, \RM{2}} = 0.02$, $D = 3$ and
    $p_{\RM{1}} = 0.5$. The black line shows the contour line for
    $J = 0$.}
  \label{J(q)_single_M7_eps002}
\end{figure}

\begin{figure}[hp]
  \centering
  \begin{subfigure}[b]{0.48\textwidth}
    \centering
    \includegraphics[width=\textwidth]{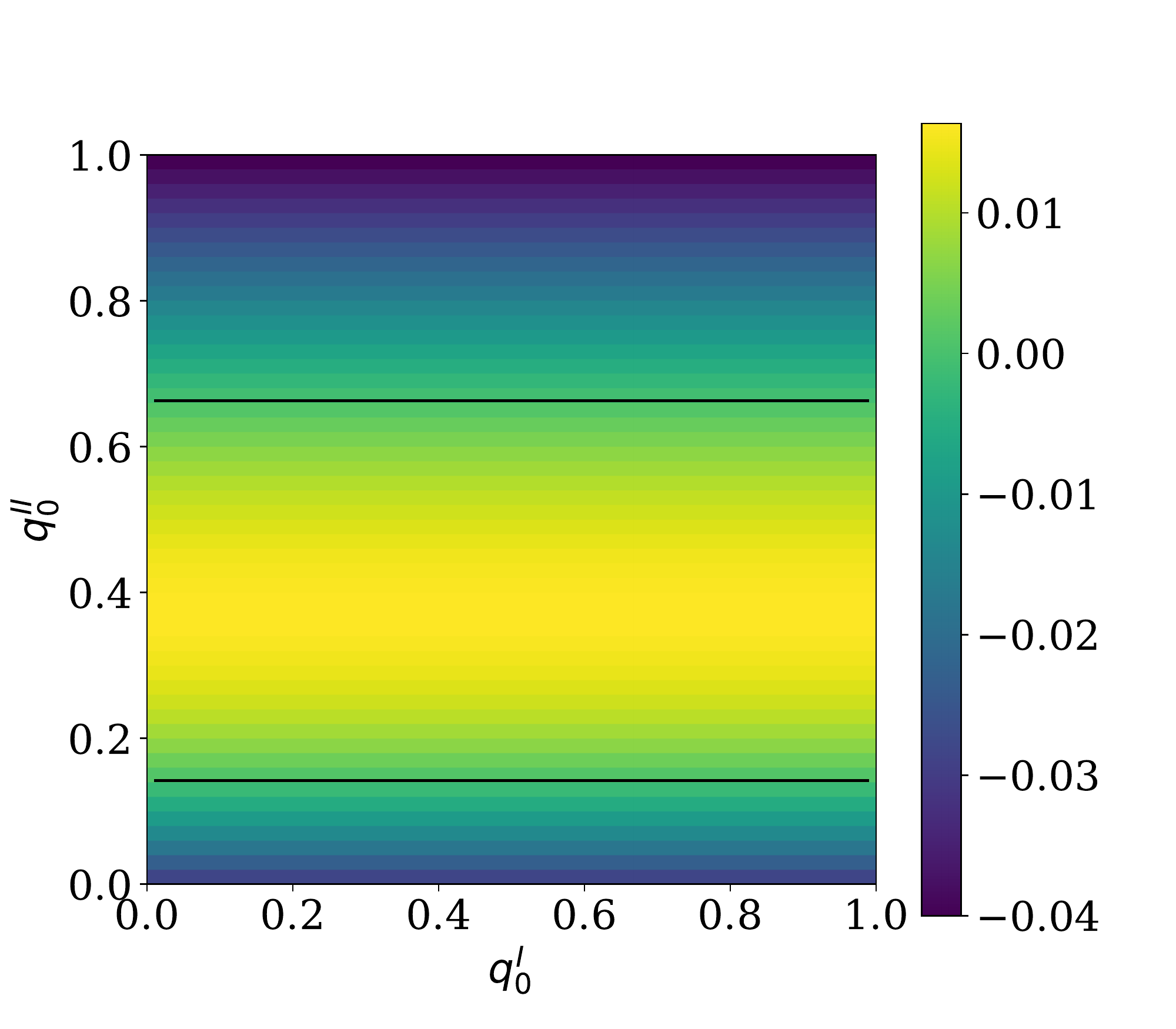}
    \caption{$p_{\RM{1}} = 0$}
    \label{J(q)_single_M5_eps002_pchoose0}
  \end{subfigure}
  \hfill
  \begin{subfigure}[b]{0.48\textwidth}
    \centering
    \includegraphics[width=\textwidth]{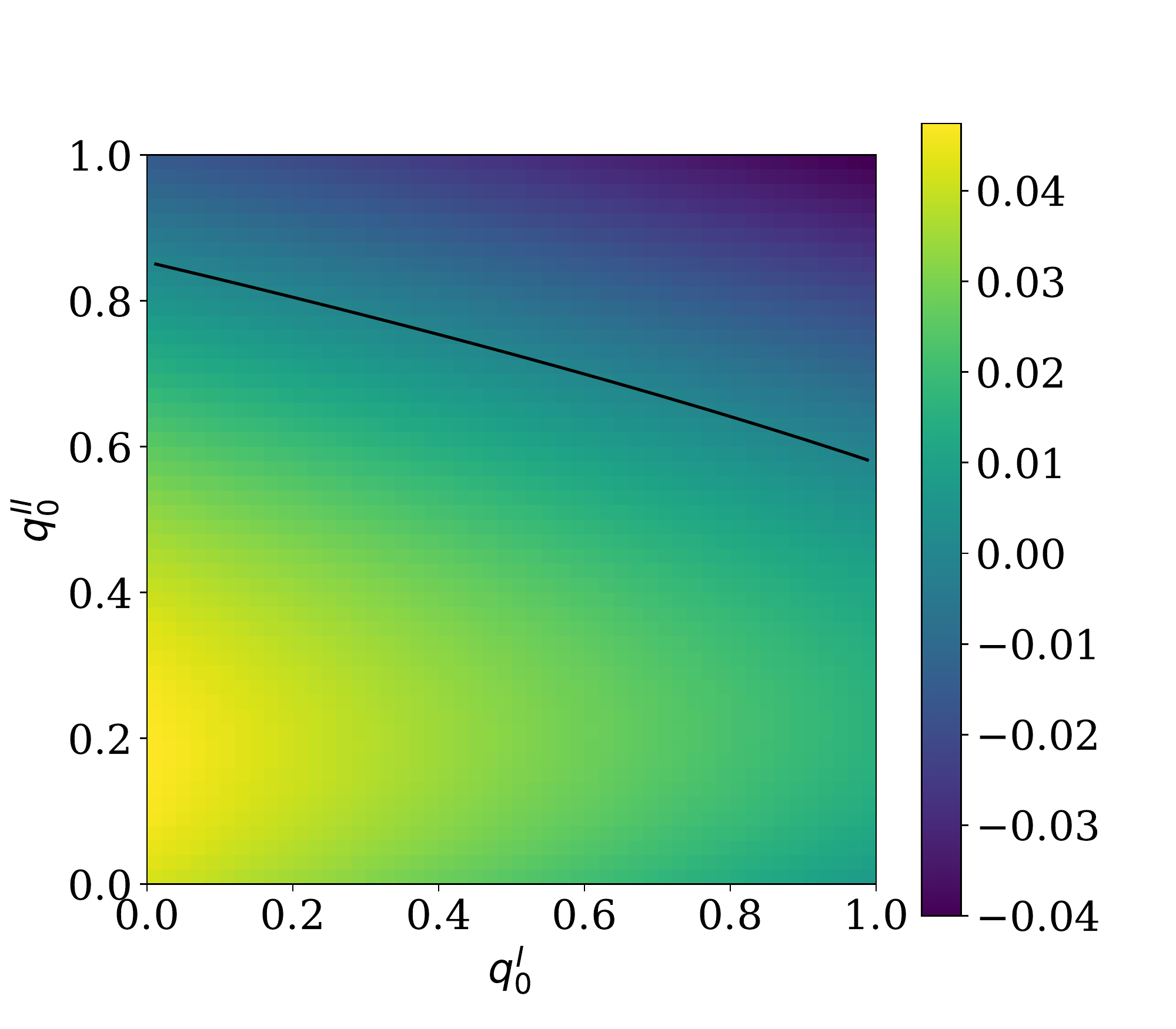}
    \caption{$p_{\RM{1}} = 0.2$}
    \label{J(q)_single_M5_eps002_pchoose02}
  \end{subfigure}
  \hfill
  \begin{subfigure}[b]{0.48\textwidth}
    \centering
    \includegraphics[width=\textwidth]{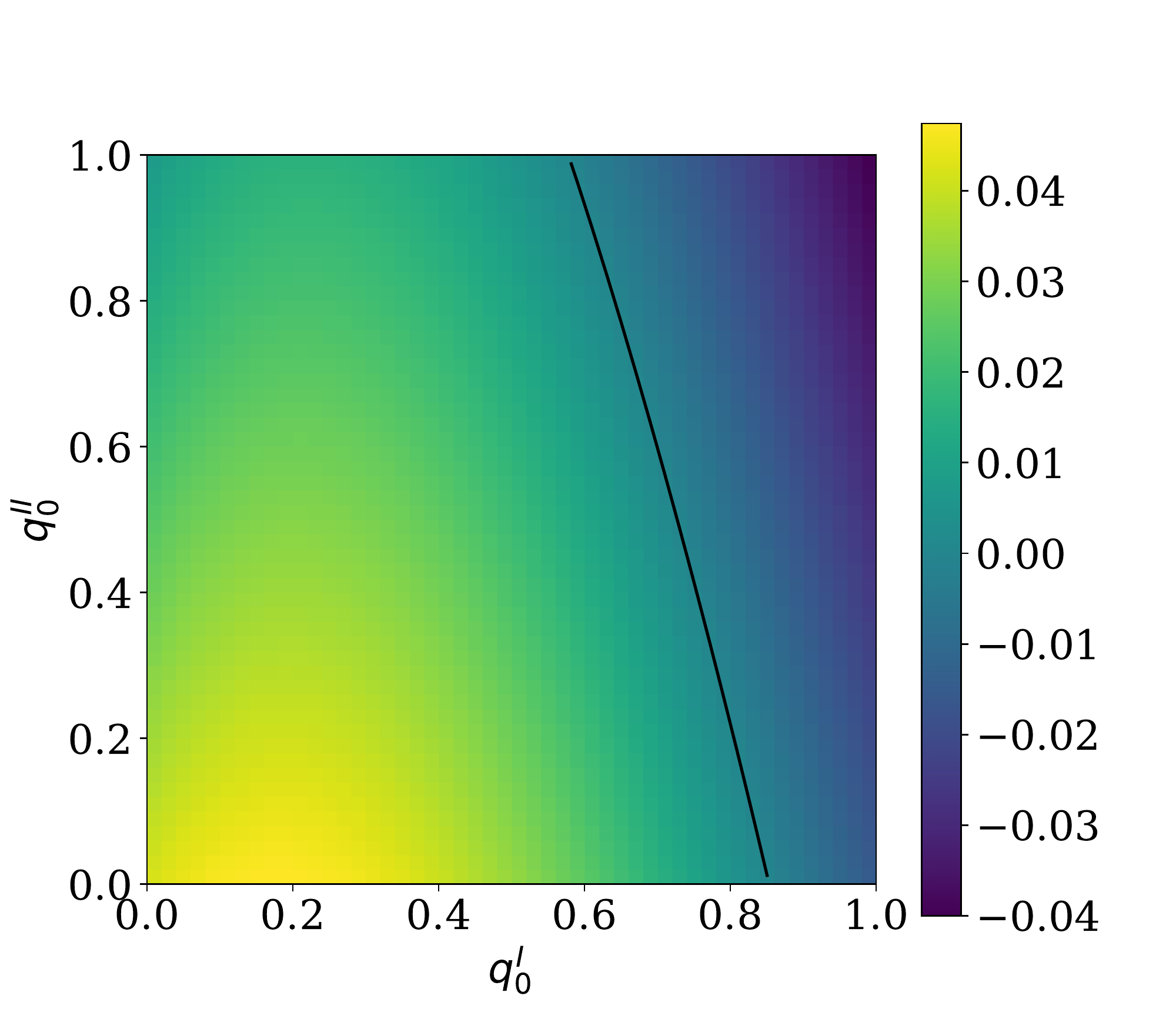}
    \caption{$p_{\RM{1}} = 0.8$}
    \label{J(q)_single_M5_eps002_pchoose08}
  \end{subfigure}
  \hfill
  \begin{subfigure}[b]{0.48\textwidth}
    \centering
    \includegraphics[width=\textwidth]{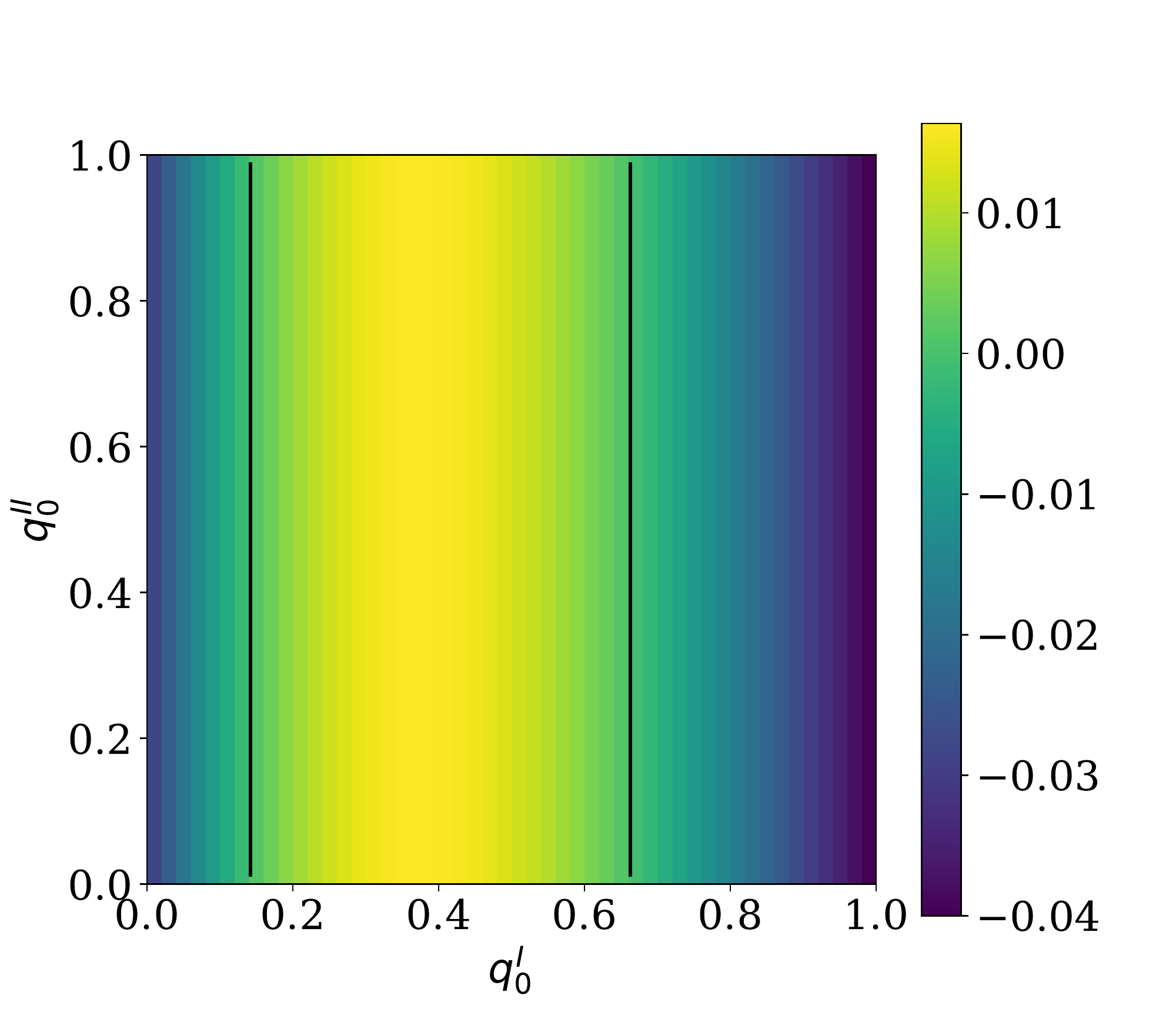}
    \caption{$p_{\RM{1}} = 1$}
    \label{J(q)_single_M5_eps002_pchoose1}
  \end{subfigure}
  \caption[Capital current of the single-play games as a function of
  $q_0^{\RM{1}, \RM{2}}$ and $p_{\RM{1}}$]{Capital current of the
    single-play games as a function of $q_0^{\RM{1}, \RM{2}}$ for
    $M = 5$, $D = 2$ and $\epsilon_{\RM{1}, \RM{2}} = 0.02$. The black
    lines reflect the contour lines for $J = 0$.}
  \label{J(q)_single_q_p}
\end{figure}

We analyse the dependence on the noise parameters. Figure
\ref{J(q)_single_M7_eps002} shows an example of the capital current as
a function of $q_0^{\RM{1}}$ and $q_0^{\RM{2}}$ for $M=7$,
$\epsilon_{\RM{1}} = \epsilon_{\RM{2}} = 0.02$, $D=3$ and
$p_{\RM{1}} = 0.5$. Furthermore, we compute the capital current as a
function of $q_0^{\RM{1}}$ and $q_0^{\RM{2}}$ for different values of
$p_{\RM{1}}$ with $M=5$, $D = 2$ and
$\epsilon_{\RM{1}, \RM{2}} = 0.02$, the results are portrayed in
figure \ref{J(q)_single_q_p}. 
  
We observe the following:

\textbf{2d) For ${p_{\RM{1}} = 0.5}$ the maximum current is reached
  for $q_0^{{\RM{1}}}=q_0^{{\RM{2}}}=0$}.  The area containing a
positive current is not finite, the current decreases with $q_0^{{\RM{1}}}$ and $q_0^{{\RM{2}}}$. This is one of the most important observations: The
maximum current is reached when both players always play game B and
there is no diffusion-like behaviour anymore. Even in game B, there is
always a chance that the other player is chosen at the beginning of a
round and the barrier is crossed more likely. However, the symmetry
within the choice of the players is important. For
$p_{\RM{1}} \neq p_{\RM{2}}$ the diffusion-like game A becomes
important again since one player is chosen more often than the other
one. This is explained in observation \textbf{2f}. For a symmetric
single-play game, the driving mechanism of the single-play coupling is
more efficient than the diffusion-like driving mechanism of the
individual games. Hence we found a coupling that dominates the
original driving mechanism!

\textbf{2e) For ${\epsilon \rightarrow 0}$ the area of positive
  current enlarges.} This is again due to the definition of the bias
parameter. It may be interesting to mention that the capital current
is always positive for $\epsilon = 0$ and $q^{\RM{1}, \RM{2}}_0 < 1$
and only vanishes for $q^{\RM{1}}_0 = q^{\RM{2}}_0 = 1$ since the
symmetric game A can not produce any directed transport.

\textbf{2f) The capital current for ${p}_{{\RM{1}}}{ = 0 (1)}$ is
  independent of ${q}_{0}^{{\RM{1}}} ({q}_{0}^{{\RM{2}}})$.}  Since
then only one player is chosen at once, this is equivalent to the
individual Parrondo games. However, the transition
$p_{\RM{1}} \in (0,1)$ is of importance which is displayed in figure
\ref{J(q)_single_q_p}. For $p_{\RM{1}} \neq 0.5$ there is an asymmetry
in the games and the single-play driving mechanism is weakened since
one player is chosen more often and therefore the diffusion-like game
A becomes more important again for this player. Hence there is some
kind of balance between the driving mechanism of the uncoupled and
single-play games which is dependent on the noise parameters.

\begin{figure}[hp]
  \centering
  \includegraphics[width=0.5\linewidth]{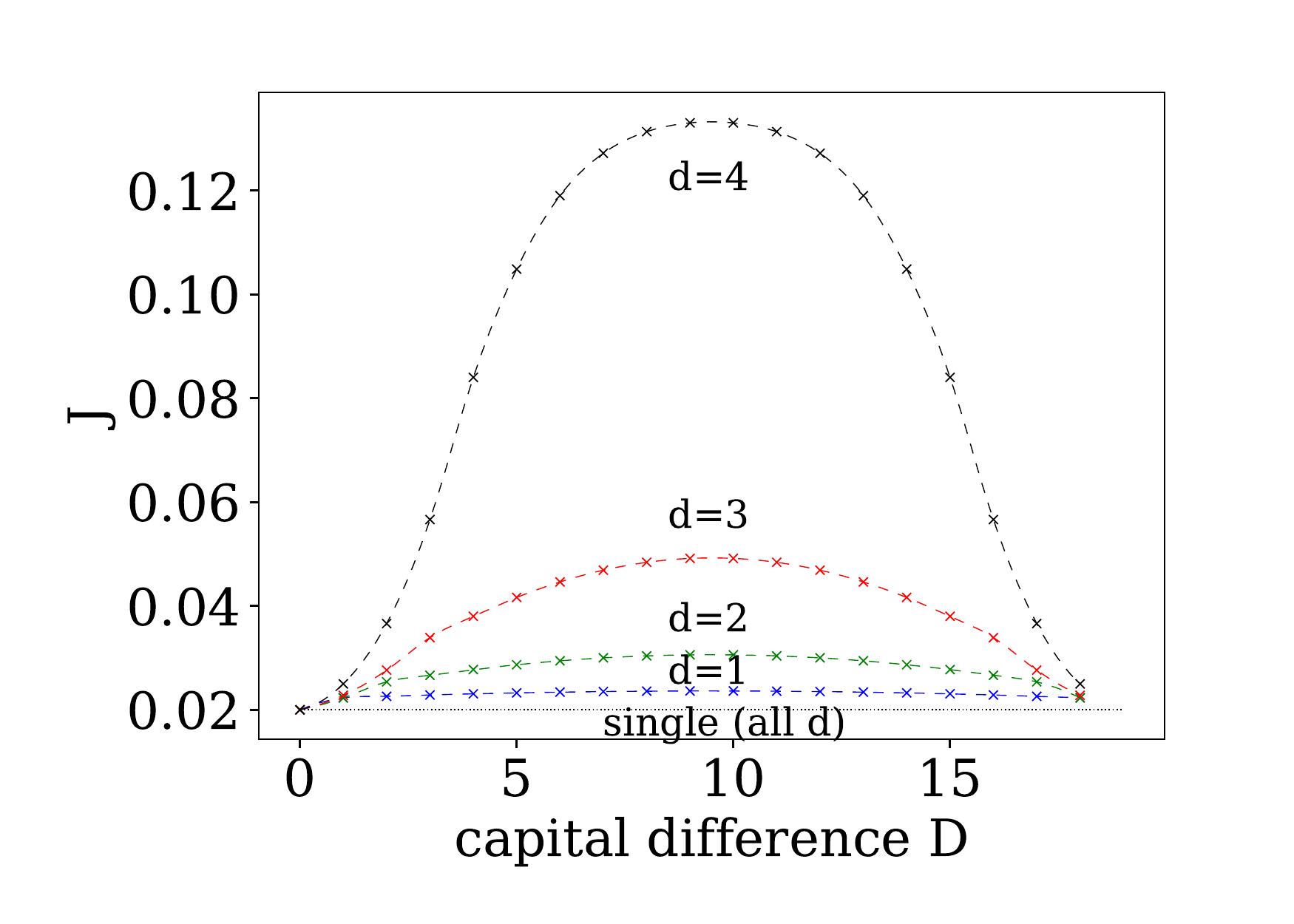}
  \caption[Capital current of the single-play games for different
  widths of the barrier]{Capital current of the single-play games for
    different widths of the barrier $d$ for $M=19$,
    $q^{\RM{1}}_0 = q^{\RM{2}}_0 = 0.5$,
    $\epsilon_{\RM{1}} = \epsilon_{\RM{2}} = 0.5 \cdot 10^{-6}$,
    $p_{\RM{1}} = 0.5$ and $q_0 = 0.5$ for the uncoupled case. The
    curve is clarified with a spline interpolation of second degree.}
  \label{J(D)_single_M19_d}
\end{figure}

We vary the width of the barrier and compute the capital current for
$M=19$, $q^{\RM{1}}_0 = q^{\RM{2}}_0 = 0.5$,
$\epsilon_{\RM{1}} = \epsilon_{\RM{2}} = 0.5 \cdot 10^{-6}$ and
$p_{\RM{1}} = 0.5$. The winning probabilities are chosen according to
table \ref{parrondo games M=19 varying d}. The results are displayed
in figure \ref{J(D)_single_M19_d}. 
  
Again, the observation \textbf{2b}
can be made, but there is also a change at the edges of the plot:

\textbf{2g) The capital current for ${D>0}$ is larger than in the
  uncoupled case. Nevertheless, the slope flattens at the edges of the
  curve.}  The single-play games differ from the double-play games in
this point: For $D=0$ the single-play games are equivalent to the
uncoupled games since the parameters are chosen to be the same for all
players and the current becomes larger for $D>0$, hence the driving
mechanism is constructive then. The reason for this might be that the
the small winning probabilities in the barrier have a larger effect
for the current when being multiplied instead of being convex combined
in the single-play coupling.

\begin{figure}[ht]
  \centering
  \includegraphics[width=0.5\textwidth]{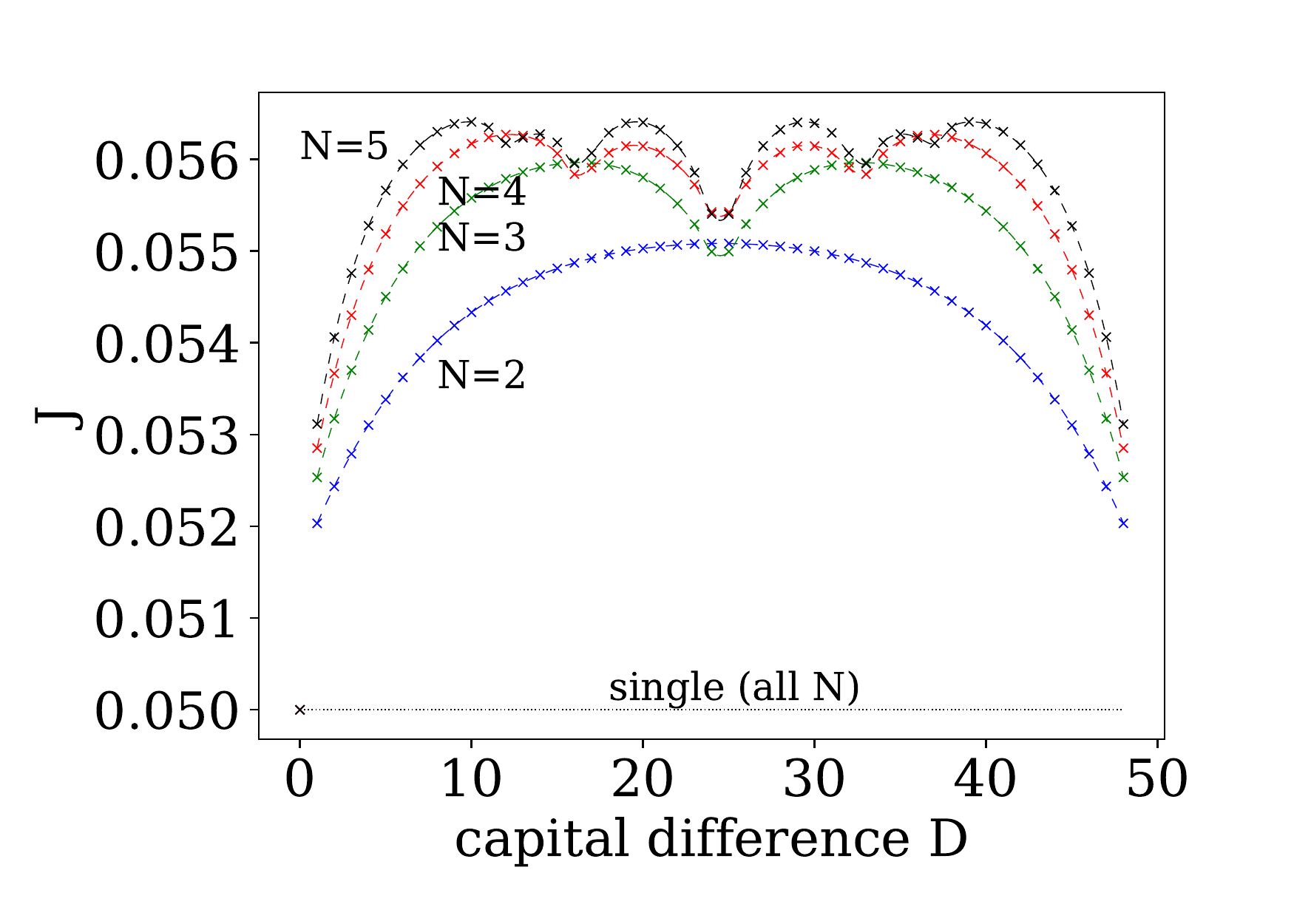}
  \caption[Capital current of the single-play games for a varying
  number of players]{Capital current of the single-play games for a
    varying number of players $N$ for $M=49$,
    $\epsilon = 0.5 \cdot 10^{-6}$,
    $p_{\RM{1}} = p_{\RM{2}} = ... = 1/N$ and $q_0 = 0.5$ for all
    players and in the uncoupled case. The curve is clarified with a
    spline interpolation of second degree.}
  \label{J(D)_single_multiplayer}
\end{figure}

We now look at multiple players for the single-play games and restrict
the discussion to equidistant capitals between consecutive players in
a certain order again. The results for $M=49$ are shown in figure
\ref{J(D)_single_multiplayer}. Many of the observations for two
players can be made. Indeed, the observations \textbf{1h} and
\textbf{1i} are also the same here. The reason is that this effect is
not dependent on the coupling but only on the distribution of capitals
over the state space.
  
\section{Discussion and Outlook}\label{Outlook}

In this paper we investigate coupled Parrondo games.
We restrict the discussion to rigid coupling and define
two different couplings: The double-play coupling assumes that both
players play the individual Parrondo games separately and have to win
or lose at the same time, respectively. This is motivated from the
rigid coupling in the continuous case. The single-play coupling, on
the other hand, is motivated by biology: At the beginning of every
round one player, whose result effects the capital of both players at
the same time, is selected randomly. In the first case the individual
probabilities are multiplied, in the second case convex combined.

The key to analyze the multi-player games is to show that for
fixed capital differences, i.e. rigid coupling, they can be reduced
to usual Parrondo games with modified parameters
and can therefore be treated in the
same way. The games converge to a stationary distribution and
this can be used to calculate the stationary current. This
allows to study the effect of the coupling and to vary the different
parameters which determine the games. 

For both couplings we show that for $M>3$ a cooperative effect
occurs. The stationary
capital current for $M>3$ is larger than in the uncoupled case, hence
the driving mechanism of the uncoupled games is supported. For the
double-play coupling the reason is the multiplication of the winning
probabilities, for the single-play coupling the choice of players at
the beginning of every round.

On the other hand, there are many differences between
the two couplings. The double-play games
show a deviation of the behaviour for $M=3$ which is not the case
in the single-play games. The change of maxima of the capital current
for different periods is positive for the double-play games but
negative for the single-play games. These and other effects
can easily be explained by the different coupling mechanism.

Apparently, the effectiveness of both couplings is dependent on the
period in different ways. The double-play coupling is more efficient
for larger, the single-play coupling for smaller periods. This always
has to be investigated in comparison to the uncoupled case which
produced almost the same current for all periods due to the modified
winning probabilities. The maximum of the capital current for $M>3$
occurs at $D \in \{\lfloor M/2 \rfloor, \lceil M/2 \rceil\}$ for both
couplings. Therefore the couplings are most efficient for the capitals
being as widely distributed over the state space as possible
(considering the symmetry).

This can be compared to the continuous case of two coupled Brownian
particles. Klumpp \emph{et al.} \cite{Klumpp_2001} determine the probability
current for a dichotomous, multiplicative noise process as a function
of the particle distance in the limit of rigid coupling,
see Fig. 3 in their paper. At first it
can be seen that the absolute value of the current
shows a maximum when the particle
difference is half the period. This is, neglecting the saturation
effects, the same behaviour as in our case. One can also observe the
symmetry between the distances $l$ and $L-l$. However, in
\cite{Klumpp_2001} an other effect becomes important: For particle
distances smaller than the width of the potential barrier, the current
hardly changes. The reason is that the coupling then can not act
over the potential barrier.

Particularly interesting is the dependence on the noise
parameters. For the double-play games the coupled
driving mechanism does not dominate but only supports the uncoupled
one: The maximum of the capital current is always found for
$q_0^{\RM{1}, \RM{2}} \notin \{0,1\}$. The single-play games show a
different result: The maximum capital current is found when both
players only play game B. The coupling dominates the noise effect of
switching the games in this case.
The reason is the new noise process, the random change between both
players. The new noise
process in the single-play coupling can dominate the original one, depending
on the choice of the respective parameters.
The double-play games do not have any additional noise
process.  The original noise process is the only one
in this case and there is no current without noise for a
vanishing bias parameter ($\epsilon=0$). 

In the end we investigate the rigid couplings for more than two
players with a constant capital difference between two consecutive
players in a certain order. The results are easy to understand: Depending on
the period and the number of players, the capital current shows
different extreme points as a function of the capital difference. We
deduce that the capital current is maximal for the capitals being as
widely distributed over the state space as possible, taking the
symmetry between the players into account.

There are many interesting questions that have not been investigated
in this work. The most interesting question is eventually
what happens if we soften the rigid coupling. A
discrete analogy of the harmonic coupling in \cite{Klumpp_2001}
is difficult to be analysed with our methods since
neither the periodic state space nor the reduction can be used. It
is possible to introduce a periodic harmonic coupling, but
such a coupling has no direct physical interpretation. The
first question one therefore needs to answer is how a
physically meaningful non-rigid coupling could look like
which pertains the periodicity.

Parrondo \emph{et al.} \cite{Parrondo_Harmer_Abbott_2000} study games which
are not capital but history dependent. Those games
can also be played by multiple coupled players. The first question
is how to introduce the coupling in that case. The second then
is whether there occur similar cooperative effects.

Another interesting aspect is the optimal choice of games and players
in every round. Dinis \cite{Dinis} examines the original Parrondo games
with Markov Decision processes, Dinis an Parrondo \cite{Dinis_Parrondo_2007}
prove that a short-range optimization of the
games can lead to a long-term loss for a positive bias parameter. The
investigation of the coupled games with these methods could lead to
interesting results, too.

Harmer \emph{et al.} \cite{Harmer_Abbott_Taylor_Parrondo_2000,
  Abbott_Harmer_Taylor_2000, Abbott_Harmer_Parrondo_Taylor_2001}
considered the recurrence and transience of the original Parrondo
games and derived conditions for the parameters leading to Parrondos
Paradox. Perhaps it may be possible to find similar conditions for the
coupled Parrondo games.

Other directions of further research are the relationship
between Parrondo games and lattice gas automata \cite{meyer-2002}
or quantum versions of Parrondo games \cite{flitney-2002-quant-parron-games}
where the effect of two or more players and the new mechanisms we found
can be present as well.
Multiple player Parrondo games thus offer a broad variety of open questions
which may be investigated in the future.

This work is based on the bachelor's thesis by Sandro Breuer.
This research did not receive any specific grant from funding
agencies in the public, commercial, or not-for-profit sectors.

\end{document}